\documentclass[12pt,a4paper]{article}  
 \usepackage[skins,theorems]{tcolorbox}

\newtcolorbox{whitebox}{colback=white,colframe=black,boxrule=0.5mm,arc=4mm,auto outer arc}

\tcbset{highlight math style={enhanced,
  colframe=red,colback=white,arc=0pt,boxrule=1pt}}
  \usepackage[bookmarksopen, bookmarksnumbered, bookmarksopenlevel=2]{hyperref}
  \usepackage{tikz}
  \usepackage{tikz-3dplot}
  \usepackage{multirow}
 \usetikzlibrary{calc}
 \usetikzlibrary{decorations} %
 \usepackage[UKenglish]{babel}
 \usepackage[toc,page]{appendix}
 \usepackage{amsmath}
 \usepackage{amssymb}
 \usepackage{graphicx}
 \usepackage{hhline}
 \usepackage[bf]{caption}
\usepackage{cite}
\usepackage[vcentermath]{youngtab}
\usepackage{geometry}
\usepackage{slashed}
\usepackage{color}
\usepackage{stackrel}
\usepackage{tikz-cd} 
\usepackage{tkz-euclide}
\usepackage{cancel} 
\usepackage{subcaption}
\usepackage[normalem]{ulem}
\usepackage{mdframed}
\usepackage{adjustbox}
\usepackage{tcolorbox}
\usepackage{enumitem}
\usepackage{empheq}
\usepackage{arydshln}
\newenvironment{eqn*}{\begin{equation*}\begin{aligned}}{\end{aligned}\end{equation*}\noindent}
\newcommand{\bqa}{\begin{eqnarray}}
\newcommand{\eqa}{\end{eqnarray}}

\hypersetup{
    pdftitle={},
    pdfauthor={},
    pdfsubject={}
}
\numberwithin{equation}{section}
\numberwithin{table}{section}

\makeatletter

\definecolor{BF}{HTML}{f903d7}

\newcommand{\be}{\begin{equation}}
\newcommand{\ee}{\end{equation}}
\newcommand{\beq}{\begin{equation}}
\newcommand{\eeq}{\end{equation}}
\newcommand{\ba}{\begin{aligned}}
\newcommand{\ea}{\end{aligned}}

\newcommand{\bea}{\begin{eqnarray}}
\newcommand{\eea}{\end{eqnarray}}

\newcommand{\cE}{\mathcal{E}}

\newcommand{\cN}{\mathcal{N}}

\newcommand{\cR}{\mathcal{R}}

\newcommand{\cV}{\mathcal{V}}

\newcommand{\cM}{\mathcal M}

\newcommand\bi{\begin{itemize}}
\newcommand\ei{\end{itemize}}

\renewcommand{\a}{{\alpha}}

\renewcommand{\l}{{\lambda}}
\def\unit{{1\kern-.65ex {\rm l}}}
\def\1{{1\kern-.65ex {\rm l}}}

\tcbset {
	base/.style={
		arc=0mm, 
		bottomtitle=0.5mm,
		boxrule=0mm,
		colbacktitle=black!10!white, 
		coltitle=black, 
		fonttitle=\bfseries, 
		left=2.5mm,
		leftrule=1mm,
		right=3.5mm,
		title={#1},
		toptitle=0.75mm,
		breakable
	}
}

\newtcolorbox{subbox}[1]{
	colframe=black!30!white,
	base={#1}
}

\newcount\hour \newcount\minute
\hour=\time \divide \hour by 60
\minute=\time
\def\now{%
\ifnum \hour<13
  \ifnum \hour=0 \advance \hour by 12 \number\hour:\else \number\hour:\fi%
     \ifnum \minute<10 0\fi%
     \number\minute%
\ A.M.%
\else \advance \hour by -12 \number\hour:%
  \ifnum \minute<10 0\fi%
  \number\minute%
  \ P.M.%
\fi%
}

\makeatother

\begin{document}

\begin{flushright}
{\tt\normalsize ZMP-HH-25/17}\\
\end{flushright}

\vskip 40 pt
\begin{center}
{\large \bf

Physics and Geometry of Complex Structure Limits \\
in Type IIB Calabi--Yau Compactifications

} 

\vskip 11 mm

Jeroen Monnee,${}^{1}$ Timo Weigand,${}^{1,2}$ and Max Wiesner${}^{1}$

\vskip 11 mm
\small ${}^{1}$\textit{II. Institut f\"ur Theoretische Physik, Universit\"at Hamburg, Notkestrasse 9,\\ 22607 Hamburg, Germany} \\[3 mm]
\small ${}^{2}$\textit{Zentrum f\"ur Mathematische Physik, Universit\"at Hamburg, Bundesstrasse 55, \\ 20146 Hamburg, Germany  }   \\[3 mm]

\end{center}

\vskip 7mm

\begin{abstract}

We provide a detailed geometric and physical interpretation of infinite distance limits in the complex structure moduli space of Type IIB compactifications on Calabi--Yau threefolds, motivated by the Emergent String Conjecture.
In the framework of semi-stable degenerations, such limits are characterised by a simple fibration structure of the fastest vanishing three-cycles.  The previously studied Hodge theoretic classification of infinite distance limits of type II, III, and IV is reflected in the number of one-cycles of the shrinking fibres. 
 Complementing our recent work on limits of type II, we focus here on type III and type IV degenerations. Based on
effective field theory considerations, these are expected to be decompactification limits to 6d and 5d, respectively.
However, establishing the existence of the associated Kaluza--Klein \emph{tower(s)} of states with both the appropriate mass scaling and the correct \emph{degeneracy} requires explicit geometric input.
   We show that the aforementioned vanishing three-cycles are special Lagrangian three-tori, thus giving rise to towers of asymptotically massless BPS particles from multi-wrapped D3-branes with the degeneracy of a Kaluza--Klein tower. We furthermore relate the BPS index of these three-cycles to the Euler characteristic of the threefold.
Finally, we systematically analyse infinite distance trajectories in multi-parameter limits described by so-called enhancement chains. We find that the primary singularity type encodes the gravitational duality frame of the limit whereas the secondary singularity type is related to the rank of the gauge group coupled to gravity. The specifics of the asymptotic physics depend crucially on whether or not the trajectory is induced by the backreaction of an EFT string.  

% \newpage 

% Een stringlimiet, zuiver en fijn, \\ 
% Kan van type II, III of IV zijn. \\
% Type II? Dan komt een snaar tevoorschijn. \\
% Type III? Zes dimensies, groot en rein.\\
% Type IV? Vijf, zo blijkt nader,\\
% Een decompactificatie zonder zolder of zolderkamer.\newline

% Neem je daarboven een extra stap,\\
% Dan toont de UV-theorie haar scherpe grap:\\
% Op species-schaal, heel precies,\\
% Bepaalt de stringlimiet het spel, als gids en als priester.\newline

% Is er geen type III of IV in zicht?\\
% Dan blijft de wereld 4D in evenwicht.\\
% Maar komt type III met zijn flair,\\
% Dan zijn we plots met z’n allen in 6D daar.\\
% Type IV, maar geen III? Let dan goed:\\
% Dan wonen we in 5D, met iets minder gemoed.\newline

% Maar pas op — als alle dijken breken,\\
% Zal Nederland de kracht der fysica zeker weten.\\
% Dan helpt geen string of dimensie meer,\\
% Dan stroomt het water, keer op keer.
\end{abstract}

\vfill

\thispagestyle{empty}
\setcounter{page}{0}
 \newpage
\tableofcontents
\vspace{25pt} 
%\newpage
\setcounter{page}{1}

\section{Introduction}
Geometry has played a key role in advancing our understanding of quantum gravity. In particular in the context of string theory compactifications, many properties of the resulting effective theories are easily accessible using the geometric features of the compactification manifold. In this way, the possible effective field theories arising from string theory can be classified in terms of the allowed geometries of the compactification manifold. String compactifications, therefore, serve as an ideal arena to delineate the landscape of effective field theories, i.e.~those EFTs arising from a UV complete theory of quantum gravity, from the Swampland consisting of EFTs that become inconsistent in the presence of gravity~\cite{Vafa:2005ui}. 

Concretely, when applied to string theory compactifications, certain Swampland constraints (see \cite{Brennan:2017rbf,Palti:2019pca,vanBeest:2021lhn,Grana:2021zvf,Agmon:2022thq} for reviews) can be translated into concrete conditions on the geometry. In this way, the Swampland programme offers an interesting avenue for studying the interplay between physics and geometry: On the one hand, known results about the geometry of compactification manifolds can be used to scrutinize our understanding of the physics of quantum gravity by explicitly testing the Swampland constraints. On the other hand, our understanding of quantum gravity can provide insights into the properties of manifolds that may not be apparent from a purely mathematical/geometric approach. 

Among the best tested Swampland constraints is the Distance Conjecture~\cite{Ooguri:2006in}, stating that in extreme limits in the moduli space of gravitational theories, there has to be a tower of states becoming exponentially light in the moduli space distance.  
 For string theory compactifications these towers of states should be detectable in the geometry of the compactification manifold. A priori, the Distance Conjecture itself does not specify what kinds of towers one should be searching for in extreme limits. Instead, the nature of the possible towers of states is believed to be constrained by the Emergent String Conjecture \cite{Lee:2019oct}. This conjecture states that at infinite distance in moduli space, the tower of states corresponds either to a KK-tower signaling a decompactification to a higher-dimensional theory or the tower of excitations of a critical string indicating that the asymptotic theory is best described by a perturbative string theory. In concrete string and M-theory settings, this pattern has been confirmed in a wide array of entire classes of compactifications~\cite{Lee:2018urn,Lee:2019oct,Lee:2019apr,Rudelius:2023odg,Lee:2021qkx,Lee:2021usk,Chen:2024cvc,Alvarez-Garcia:2023gdd,Alvarez-Garcia:2023qqj,Alvarez-Garcia:2021pxo,Marchesano:2019ifh,Baume:2019sry,Blumenhagen:2023yws,Xu:2020nlh,Lee:2019tst,Klaewer:2020lfg,Basile:2022zee,Etheredge:2023odp,Aoufia:2024awo,Calderon-Infante:2024oed,Friedrich:2025gvs,Gkountoumis:2025btc}, oftentimes crucially relying on the geometric properties of the compactification manifold. Complementary to these approaches, there have been numerous efforts to motivate the Emergent String Conjecture from the bottom-up using arguments based on black hole thermodynamics \cite{Basile:2023blg,Basile:2024dqq,Herraez:2024kux}, the species entropy \cite{Cribiori:2023ffn}, gravitational scattering amplitudes \cite{Bedroya:2024ubj}, or by employing the consistency of supergravity strings in 5d $\mathcal{N}=1$ theories \cite{Kaufmann:2024gqo}. This bottom-up evidence suggests that the Emergent String Conjecture indeed captures universal features of quantum gravity and can thus shed new light onto the features of the geometries arising in asymptotic regimes of string theory compactifications. 

In this work, we continue our study, initiated in \cite{Friedrich:2025gvs}, of the physical and geometric interplay of infinite-distance limits in the vector multiplet moduli space of Type IIB string theory compactified on Calabi--Yau threefolds, in light of the 
 Emergent String Conjecture. 
 In combination, \cite{Friedrich:2025gvs} and the present work establish the existence of towers of BPS particles becoming light based on a detailed understanding of the asymptotic geometry in complex structure moduli space, and 
 provide evidence that the BPS indices characterising the degeneracies of these towers are as expected for the states to play the role of Kaluza-Klein or emergent string states in the sense of \cite{Lee:2019oct}.  

From a geometric point of view, the moduli residing in the vector multiplets of the four-dimensional $\mathcal{N}=2$ supergravity theory describe the complex structure deformations of the underlying Calabi--Yau threefold. These, in turn, can be thought of as parametrising the volumes of certain 3-cycles in the threefold. In particular, in infinite-distance limits the complex scalar fields associated with the complex structure deformations can be split into an axionic part and a saxionic part. The infinite distance limit then corresponds to scaling the saxionic part to infinity. Geometrically, in such limits various legs of such 3-cycles grow or shrink, 
 leading to an extremely anisotropic threefold. The precise rates at which the various 3-cycles may grow/shrink are tightly constrained by the underlying limiting mixed Hodge structures \cite{schmid,CKS,Grimm:2018ohb,Grimm:2018cpv,Grimm:2019wtx,Gendler:2020dfp,Bastian:2020egp,Calderon-Infante:2020dhm,Grimm:2021ikg,Palti:2021ubp,Grimm:2021vpn,Grimm:2022xmj,Bastian:2023shf,vandeHeisteeg:2022gsp,Monnee:2024gsq} that describe the breakdown of the cohomology of the Calabi--Yau threefold in the limit. The different possibilities may be broadly classified into the principal types II, III, or IV, which are each expected to lead to different emergent physics. 

In recent work \cite{Friedrich:2025gvs}, the type II limits that are realised in so-called Tyurin degenerations were found to correspond to emergent string limits in which the emergent dual frame is described by a heterotic string with target space ${\rm K3}\times T^2$, in accordance with the expectations from the Emergent String Conjecture. In this work, we instead turn our attention to the interpretation of type III and type IV limits, which are expected to correspond to decompactification limits to 6d and 5d, respectively. One motivation for this
 stems from the interpretation of the mirror dual Type IIA string theory as the corresponding decompactification limits \cite{Lee:2019oct}, even though this logic only applies to limits close to the large complex structure / large volume point.
 Hence, at least in the large complex structure / large volume regimes, the behaviour of the low-energy effective physics is expected to be the same. 

It has been suggested that the objects in the 4d $\mathcal{N}=2$ supergravity theory that may give rise to the expected KK towers are $\frac{1}{2}$-BPS particles which geometrically arise from D3-branes wrapped on special Lagrangian 3-cycles inside the Calabi--Yau threefold \cite{Grimm:2018ohb}. For every infinite-distance limit in the vector multiplet moduli space, a subset of these states indeed becomes asymptotically massless and weakly coupled as dictated by the growth theorems of \cite{schmid,CKS}. More generally, the structure of the low-energy effective theory is well understood using techniques from asymptotic Hodge theory \cite{Grimm:2018ohb,Grimm:2018cpv,Grimm:2019wtx,Gendler:2020dfp,Bastian:2020egp,Calderon-Infante:2020dhm,Grimm:2021ikg,Palti:2021ubp,Grimm:2021vpn,Grimm:2022xmj,Bastian:2023shf}, see also \cite{vandeHeisteeg:2022gsp,Monnee:2024gsq} for reviews and further references. In addition, along trajectories in the moduli space induced by the backreaction of so-called EFT strings, which are $\frac{1}{2}$-BPS solitonic string solutions in the 4d theory, there appear to be candidate states whose mass-scale exhibits the expected scaling corresponding to a tower of KK-states \cite{Lanza:2020qmt,Lanza:2021udy}. However, it has not been shown whether there indeed exists a \emph{tower} of such states. Complementary to this, it has been argued purely in terms of algebraic properties of the low-energy effective theory that for some of the infinite-distance limits there can, in fact, be a tower of asymptotically massless states by employing the action of the local monodromy \cite{Grimm:2018ohb,Grimm:2018cpv,Corvilain:2018lgw}. However, these towers do \emph{not} have the correct mass-scale that would be expected for a KK tower corresponding to a decompactification to 5d/6d. For example, in the large complex structure limit the monodromy towers necessarily involve bound states of D0/D2-branes in Type IIA language, whereas the leading tower of states corresponds to the pure D0-brane tower, signaling a decompactification to 5d M-theory. Another point worth emphasising is that from a purely algebraic perspective it is not clear how to address the \emph{degeneracy} of states in the tower, which is a crucial ingredient in order to establish the microscopic interpretation of the tower as a KK tower. 

In this work, we overcome these challenges by supplementing the aforementioned algebraic perspective with a geometric characterisation of the infinite-distance limits, building on the recent work \cite{Friedrich:2025gvs} as well as similar works \cite{Lee:2021qkx,Lee:2021usk,Alvarez-Garcia:2023gdd,Alvarez-Garcia:2023qqj} which concern complex structure limits of Weierstrass models in the context of F-theory compactifications. 
 We phrase the discussion in the framework of a
 semi-stable degeneration of the Calabi--Yau threefold. In these limits, the Calabi--Yau splits into various components that can intersect over surfaces, curves, and points. Whether a limit is of type II, III, or IV is then straightforwardly interpreted in terms of the intersection/dual-graph of the singular threefold. Importantly,
 one can completely characterise the cohomology of the singular threefold in terms of a ``geometric'' mixed Hodge structure due to Deligne \cite{Deligne:1971,Deligne:1974}, which can furthermore be matched with the limiting mixed Hodge structure that describes the asymptotic behaviour of the couplings in the low-energy effective theory. Using this input, we are indeed able to establish the physical interpretation of type III/IV limits as decompactification limits to 6d/5d.  Turning tables around, based on this physical input, we are furthermore able to give a prediction for the BPS index counting of the associated A-branes wrapping certain special Lagrangian cycles in the threefold. In addition, we also study the interpretation of multi-parameter limits described by so-called enhancement chains, and highlight which part of the chain is relevant in order to establish the emergent gravitational theory for different classes of trajectories in the moduli space. In the following, we list the main results obtained in this paper. 

\subsection*{Summary of Results}

\paragraph{Tower(s) of BPS states.}
By leveraging the relation between the geometric mixed Hodge structure associated with the singular threefold and the limiting mixed Hodge structure near a semi-stable degeneration, we argue for the existence of special Lagrangian 3-cycles $\Gamma$ that locally admit a torus-fibration of the form
\begin{equation} \label{Gamm-def1}
    T^d\hookrightarrow \Gamma\to B_{3-d}\,,
\end{equation}
where $d=1,2,3$ for limits of type II/III/IV, respectively. The base $B_{3-d}$ is thus respectively a curve, a circle or a point in these three cases. Importantly, upon approaching the infinite-distance limit the $T^d$-fibre shrinks to a point, such that D3-branes wrapping the 3-cycle give rise to asymptotically massless and weakly-coupled BPS states. Crucially, for $d=2,3$ the topology of the 3-cycle trivially allows for multi-wrappings, such that one in fact obtains a \emph{tower} of light BPS states.\footnote{The case $d=1$ was already discussed in detail in \cite{Friedrich:2025gvs}, and towers from multi-wrapped D3-branes were found for those 3-cycles whose base $B_{2}$ has a non-negative intersection number in a K3 or $T^4$ appearing at the intersection of the components of the threefold in the stable degeneration.} Furthermore, by analysing the degeneracy of states via the BPS index, we find that the towers ought to be interpreted as KK towers signaling a decompactification limit. The precise results for the type III/IV limits are stated below.  

\vspace{2mm}
\begin{mdframed}[backgroundcolor=white,shadow=true,shadowsize=4pt,shadowcolor=black,roundcorner=6pt]
\textbf{Type III limits:}\\
In Type IIB string theory compactified on Calabi--Yau threefolds, there are precisely two leading towers of BPS particles that become light at the same rate along any semi-stable degeneration realising a type III limit in the complex structure moduli space of the threefold. Along trajectories induced by the backreaction of EFT strings, the mass of the BPS particles scales as
\begin{equation}
    \frac{m}{M_{\rm Pl}}\sim e^{-\gamma \Delta}\,,\qquad \gamma=1\,,
\end{equation}
where $\Delta$ denotes the distance with respect to the moduli space metric. In addition, the BPS indices are given by
\begin{equation}
    \Omega_{\rm BPS}(n\Gamma^{(i)}) = -\chi(V)\,,\qquad \forall n\in\mathbb{Z}_{>0}\,,  \qquad i = A,B \,,
\end{equation}
where $\Gamma^{(i)}$, $i=A,B$, denote the two special Lagrangian 3-cycles of the form (\ref{Gamm-def1}) with $d=2$ appearing in the limit that are wrapped by the D3-branes, and $\chi(V)$ denotes the Euler character of the Calabi--Yau threefold. The physical interpretation of the type III EFT string limit is a decompactification limit to 6d.
\end{mdframed}

\vspace{2mm}
\begin{mdframed}[backgroundcolor=white,shadow=true,shadowsize=4pt,shadowcolor=black,roundcorner=6pt]
\textbf{Type IV limits:}\\
In Type IIB string theory compactified on Calabi--Yau threefolds, there is precisely one leading tower of BPS particles becoming light along any semi-stable degeneration realising a type IV limit in the complex structure moduli space of the threefold. Along trajectories induced by the backreaction of EFT strings, the mass of the BPS particles scales as
\begin{equation}
    \frac{m}{M_{\rm Pl}}\sim e^{-\gamma \Delta}\,,\qquad \gamma=\sqrt{\frac{3}{2}}\,,
\end{equation}
where $\Delta$ denotes the distance with respect to the moduli space metric. In addition, the BPS indices are given by
\begin{equation}
    \Omega_{\rm BPS}(n\Gamma) = -\chi(V)\,,\qquad \forall n\in\mathbb{Z}_{>0}\,,
\end{equation}
where $\Gamma$ denotes the special Lagrangian 3-cycle of the form (\ref{Gamm-def1}) with $d=3$ that is wrapped by the D3-brane, and $\chi(V)$ denotes the Euler character of the Calabi--Yau threefold. The physical interpretation of the type IV EFT string limit is a decompactification limit to 5d.
\end{mdframed}

\paragraph{Enhancements.}
For limits involving multiple moduli, the emergent physics will depend crucially on both the number of moduli that are sent to infinity, as well as the hierarchy between them. We find that it is important to distinguish between trajectories which are induced by the backreaction of EFT strings -- such paths are linear in the saxions -- and those that are not.\footnote{The distinguished behaviour of linear paths in the moduli space and their relevance with respect to the Distance Conjecture was also emphasised in \cite{Grimm:2022sbl}.} In the language of asymptotic Hodge theory, the different singularity types arising in multi-parameter limits are encoded in terms of an enhancement chain of limiting mixed Hodge structures. Importantly, across the enhancement chain the primary singularity type (II/III/IV) may change, as well as the secondary singularity type typically denoted by a subscript (II$_b$, III$_c$, IV$_d$). We find that the first possibility is associated with a change in the \emph{gravitational} features of the theory, whereas the latter is related to the rank of the gauge group of decoupled field theory sectors. The latter statement is in agreement with the analysis~\cite{Marchesano:2023thx,FierroCota:2023bsp,Marchesano:2024tod,Castellano:2024gwi} of rigid field theory subsectors in the mirror dual Type IIA setup. For simplicity, we summarise the main result in the case of 2-parameter limits, and refer the reader to Sections \ref{sec:EFT-POV} and  \ref{sec_Enhancements} for further details and explanation of the notation. \newline

\begin{mdframed}[backgroundcolor=white,shadow=true,shadowsize=4pt,shadowcolor=black,roundcorner=6pt]
\textbf{Physics of enhancement chains:}\\
Consider an enhancement chain of the form
\begin{equation}
    \text{type}_{a_1}(\Delta_{k_1})\to \text{type}_{a_2}(\Delta_{k_1k_2})\,,
\end{equation}
where type$_{a_r}(\Delta_{k_1\cdots k_r})$ denotes the singularity type along a codimension-$r$ singularity in the moduli space, including the secondary singularity type denoted by $a_r$. Then we distinguish two cases:
\begin{itemize}
    \item \textbf{Case (A): non-EFT string limit}\\
    If there is an asymptotic hierarchy between two saxions $s^{k_1}\gg s^{k_2}$, then the primary singularity type of the \emph{first} singularity in the chain, i.e.~at $\Delta_{k_1}$, determines the quantum gravitational duality frame emerging in this limit. 
    The subsequent enhancements of the primary singularity type indicate that within this duality frame an additional limit is taken. 
    \item \textbf{Case (B): EFT string limit}\\
    If the two saxions scale at the same rate $s^{k_1}\sim s^{k_2}$, then it is the primary singularity type of the \emph{second} singularity in the chain, i.e. at $\Delta_{k_1,k_2}$,
     that determines the quantum gravitational duality frame emerging in this limit. In addition, the change in the secondary singularity type determines the number of $\mathrm{U}(1)$ gauge fields that are decoupled from gravity for the singularity of $\text{type}_{a_1}(\Delta_{k_1})$, but get recoupled to gravity in the EFT string limit.
\end{itemize}

\end{mdframed}

\subsection*{Structure of the Paper}
The article is structured as follows. In Section \ref{sec:EFT-POV} we give the algebraic description of type II/III/IV limits in the complex structure moduli space of a Calabi--Yau threefold, and characterise the scaling of the masses of BPS particles and tensions of EFT strings using asymptotic Hodge theory. In Section \ref{sec:II_III_IV} we provide the geometric interpretation of type II/III/IV limits 
 in the framework of semi-stable degenerations. Importantly, we review the relation between the algebraic limiting mixed Hodge structure used in Section \ref{sec:EFT-POV} and the geometric mixed Hodge structure that carries information about the cohomology of the singular threefold. In Section \ref{sec:decomp} we then use this relation to obtain the appropriate shrinking 3-cycles and establish the type III/IV limits as decompactification limits to 6d/5d. In Section \ref{sec_Enhancements} we discuss the geometric and physical interpretation of enhancement chains of singularities, in particular along trajectories that are not EFT string limits. Finally, Appendix \ref{app:dual-graph} contains additional background on the differentio-geometric interpretation of the dual graph of the semi-stable degeneration.

\section{EFT Perspective on Complex Structure Degenerations}
\label{sec:EFT-POV}
In this section, we discuss the basis for studying infinite-distance limits in the vector multiplet moduli space of four-dimensional $\mathcal{N}=2$ supergravity theories that arise as the low-energy effective description of Type IIB string theory compactified on Calabi--Yau threefolds. In particular, we briefly recall the classification of such limits into type II/III/IV in terms of the underlying web of limiting mixed Hodge structures \cite{Grimm:2018cpv,Grimm:2018ohb}.

To fix our notation, recall that the low-energy effective description of Type IIB string theory compactified on a Calabi--Yau threefold $V$ is a four-dimensional $\mathcal{N}=2$ supergravity theory with a single gravity multiplet, $h^{2,1}(V)$ vector multiplets, and $h^{1,1}(V)+1$ hypermultiplets. Focusing on the gravity and vector multiplet sector, the relevant bosonic part of the four-dimensional action reads
\begin{equation}
\label{eq:action_4d}
    S_4 = \int \frac{1}{2}M_{\mathrm{Pl}}^2R\star1 +G_{i\bar{\jmath}}\,\mathrm{d}u^i\wedge\star\,\mathrm{d}u^{\bar{\jmath}}+\frac{1}{4}\mathcal{I}_{IJ}F^I\wedge\star\,F^J+\frac{1}{4}\mathcal{R}_{IJ}F^I\wedge F^J\,.
\end{equation}
Here $u^i$, for $i=1,\ldots, h^{2,1}(V)$, denote the complex structure moduli of the Calabi--Yau threefold $V$, which make up the complex scalars in the vector multiplets, and $F^I=\mathrm{d}A^I$, where $A^I$, for $I=0\ldots, h^{2,1}(V)$, denote the $\mathrm{U}(1)$ gauge fields residing in the gravity and vector multiplets. Instead of the local coordinates $u^i$, it is useful to define the covering coordinate 
\begin{equation}
    t^i = a^i + i s^i:= \frac{1}{2\pi i} \log u^i\,,
\end{equation}
where the $a^i$ are referred to as axionic and the $s^i$ as saxionic components of the complex fields $t^i$. 

Of primary interest is the metric $G_{i\bar{\jmath}}$ on the complex structure moduli space, which may locally be expressed in terms of a K\"ahler potential
\begin{equation}\label{eq:metric}
    G_{i\bar{\jmath}} = \partial_i \bar{\partial}_{\bar{\jmath}}K\,,\qquad K=-\log\|\Omega\|^2\,,
\end{equation}
where $\Omega$ denotes the unique holomorphic $(3,0)$-form on the Calabi--Yau threefold $V$, and we have introduced the notation
\begin{equation}
    \langle v,w\rangle = \int_V v\wedge\star\,\bar{w}\,,\qquad \|v\|^2 = \langle v, v\rangle\,,\qquad v,w\in H^3(V,\mathbb{C})\,,
\end{equation}
for the \emph{Hodge inner product} and \emph{Hodge norm} on $H^3(V,\mathbb{C})$, respectively. 
There are two kinds of light states to consider. 
\begin{itemize}
    \item \textbf{Particles States:}\\
    In the present setup, there can exist particles that are charged under the $\mathrm{U}(1)$ gauge fields $A^\alpha$ obtained from the reduction of the RR four-form field
    \begin{equation}
        C_4 = A^\alpha\wedge\gamma_\alpha\,,\qquad \alpha=1,\ldots, 2h^{2,1}+2\,.
    \end{equation}
    Such particles are specified by a choice of charge vector $q\in H^3(V,\mathbb{Z})$. Expanding $q=q^\alpha\gamma_\alpha$ with respect to an integral basis $\{\gamma_\alpha\}$ of $H^3(V,\mathbb{Z})$, the coefficients $q^\alpha$ precisely correspond to the $\mathrm{U}(1)$ charges. By the BPS condition, the mass $m_q$ is bounded from below by the central charge $\mathcal{Z}_q$ as
    \begin{equation}
    \label{eq:BPS-mass}
        \frac{m_q}{M_{\mathrm{Pl}}} \geq |\mathcal{Z}_q| = \frac{|\langle q,\Omega\rangle|}{\|\Omega\|}\,,
    \end{equation}
    which is saturated if the particle state is BPS. In addition, the physical charge $\mathcal{Q}_q$ is given by
    \begin{equation}
    \label{eq:physical-charge}
        \mathcal{Q}_q = \|q\|\,,
    \end{equation}
    such that the asymptotically weakly-coupled particles are those for which $\|q\|\to 0$. 
    
    \item \textbf{EFT strings:}\\
    Besides the particle states, the theory also contains so-called EFT strings. These arise as solitonic $\frac{1}{2}$-BPS solutions to the equations of motion of the four-dimensional supergravity theory \cite{Greene:1989ya,Lanza:2020qmt,Lanza:2021udy}. Importantly, such strings are magnetically charged under the axions $a^i = \mathrm{Re}\,t^i$ and thus induce a backreaction on the complex stucture moduli of the form
    \begin{equation}
    \label{eq:EFT-string-backreaction}
        t^i = \frac{e^i}{2\pi i} \log\left(\frac{z}{z_0}\right)\,.
    \end{equation}
    Here $z$ denotes the coordinate transverse to the EFT string in the four-dimensional spacetime, with the string core located at $z=z_0$, and $e^i$ denotes the magnetic charges of the EFT string. The tension of the EFT string is straightforwardly expressed in terms of its magnetic charges and the K\"ahler potential via\footnote{To be precise, one first has to consider a loop-configuration of the string in order to regulate its backreaction at infinity. Additionally, the derivation of \eqref{eq:EFT-string-tension} crucially relies on the existence of an axionic shift symmetry $t^i\mapsto t^i+c^i$ of the K\"ahler potential, which is known to hold asymptotically. }
    \begin{equation}
    \label{eq:EFT-string-tension}
        \frac{T_e}{M_{\text{Pl}}^2} = - e^i \frac{\partial K}{\partial s^i}\,.
    \end{equation}
\end{itemize}
From the relations \eqref{eq:BPS-mass}, \eqref{eq:physical-charge}, and \eqref{eq:EFT-string-tension}, it is clear that understanding the scaling of the Hodge norm/inner product is essential in order to study which states become light, and at what rate. We are interested in extreme limits corresponding to singular loci $\Delta_{k_1,\dots, k_n}\subset\cM_{\rm c.s.}(V)$ defined in terms of the local coordinates $u^k$ as 
\begin{equation}
    \Delta_{k_1,\dots,k_n}=\{u^{k_1}=\dots = u^{k_n}=0\} \,. 
\end{equation}
In the vicinity of such a singular locus, the scalings of the Hodge norm/inner product can be computed using the limiting mixed Hodge structure on $H^3(V,\mathbb{C})$. While for further details we refer to \cite{Grimm:2018ohb,Grimm:2018cpv,Grimm:2019wtx,Gendler:2020dfp,Bastian:2020egp,Calderon-Infante:2020dhm,Grimm:2021ikg,Palti:2021ubp,Grimm:2021vpn,Grimm:2022xmj,Bastian:2023shf,vandeHeisteeg:2022gsp,Monnee:2024gsq} and references therein, for our applications, it suffices to know that such a mixed Hodge structure is determined by a splitting of $ H^3(V,\mathbb{C})$ as 
\begin{equation}
    H^3(V,\mathbb{C}) = \bigoplus_{0\leq p,q\leq 3} I^{p,q}(\Delta_{k_1,\dots,k_n})\,. 
\end{equation}
The properties of the singularity are then encoded in the dimensions 
\begin{equation}
    i^{p,q} = \text{dim}\,I^{p,q}(\Delta_{k_1,\dots, k_n})\,. 
\end{equation}
Due to the Calabi--Yau condition for $V$ only one $i^{3,q}$ for $q=0,\dots, 3$ can be non-zero, and correspondingly the limits are classified into type I ($q=0$), II ($q=1$), III ($q=2$) or IV ($q=3$). Due to symmetry relations between the various spaces $I^{p,q}$, there only remains a single free parameter, which can be chosen to be $i^{2,2}$. This results in a classification of the limits according to their primary type (I, II, III, IV) and their secondary type $i^{2,2}$, typically denoted as a subscript.

To compute the scaling of the Hodge norm in the vicinity of $\Delta_{k_1,\dots, k_n}$, we can consider the graded spaces 
\begin{equation}
    \text{Gr}_{\ell_n}(\Delta_{k_1,\dots, k_n}) = \bigoplus_{p+q=\ell_n} I^{p,q}(\Delta_{k_1,\dots, k_n})\,. 
\end{equation}
Suppose we approach the locus $\Delta_{k_1,\dots,k_n}$ along a growth sector 
\begin{equation}
    \cR_{k_1,\dots, k_n}= \left\{t^j\Bigg| \frac{s^{k_1}}{s^{k_2}}\,,\dots \,, \frac{s^{k_{n-1}}}{s^{k_n}}, s^{k_n}>\gamma\right\}\,,\qquad \gamma>1\,.
\end{equation}
For a path approaching $\Delta_{k_1,\dots, k_n}$ inside $\cR_{k_1,\dots,k_n}$, the growth theorem~\cite{schmid,CKS} tells us that the Hodge norm for an element $q\in \text{Gr}_{\ell_1}(\Delta_{k_1})\cap \text{Gr}_{\ell_2}(\Delta_{k_1,k_2})\cap\cdots \cap \text{Gr}_{\ell_n}(\Delta_{k_1,\dots,k_n})$ asymptotically scales as  
\begin{equation}\label{eq:growththeorem}
    \|q\|^2 \sim \prod_{i=1}^{n}\left(\frac{s^{k_i}}{s^{k_{i+1}}}\right)^{\ell_i-3}\,. 
\end{equation}
The asymptotic scaling of the K\"ahler potential can similarly be determined: Define an integer $d_{k_1\cdots k_i}$ such that $d_{k_1\cdots k_i}=0,1,2,3$ if the limiting mixed Hodge structure associated with $\Delta_{k_1\cdots k_i}$ is of (primary) type I, II, III, IV, respectively. Then
\begin{equation}\label{Kgrowththeorem}
   e^{-K}= \|\Omega\|^2\sim \prod_{i=1}^n \left(\frac{s^{k_i}}{s^{k_{i+1}}}\right)^{d_{k_1\cdots k_i}}\,.
\end{equation}

Finally, note that the smallest value of $\ell_i$ for which ${\rm Gr}_{\ell_i}(\Delta_{k_1\cdots k_i})$ is not empty is given by $\ell_i = 3 - d_{k_1 ,\ldots, k_i}$.
For endpoints of type II, III, IV, these are the spaces ${\rm Gr}_2(\Delta_{k_1\cdots k_i})$, ${\rm Gr}_1(\Delta_{k_1\cdots k_i})$ and ${\rm Gr}_0(\Delta_{k_1\cdots k_i})$, respectively.
 As we will review below,  this is very important because the elements of these spaces correspond to charges of particles whose mass generically vanishes at the fastest rate in the limit under consideration.

\subsubsection*{EFT Characterisation of II/III/IV}
In light of the Distance Conjecture and the Emergent String Conjecture~\cite{Ooguri:2006in,Lee:2019oct}, we are interested in studying how the mass scale of the lightest tower of states behaves along an infinite-distance path in the complex structure moduli space. For simplicity, let us consider a \emph{linear} path that approaches the singular divisor $\Delta_{k_1\cdots k_n}$, parametrised by 
\begin{equation}
\label{eq:path}
    s^{k_i} = e^{k_i} \lambda\,,\qquad i=1,\ldots, n\,,
\end{equation}
for some positive non-zero constants $e^{k_i}$, while the remaining saxions are kept finite in the limit $\lambda\to\infty$. Such a homogeneous limit (in $\lambda$) is precisely the one induced by the backreaction of an EFT string with charges $\{e^{k_i}\}$, recall equation \eqref{eq:EFT-string-backreaction}.\footnote{It has been suggested in \cite{Grimm:2022sbl} that, in general, the issues of path-dependence in multi-parameter models may in fact be reduced to such linear paths by employing the tameness of the underlying functions. Let us further remark that the Distance Conjecture is supposed to apply to \emph{geodesics}. The linear paths \eqref{eq:path} typically correspond to geodesics with respect to the moduli space metric, at least asymptotically, see for example \cite{Calderon-Infante:2022nxb} for an in-depth discussion on this.}
Therefore, we will refer to trajectories of the form \eqref{eq:path} as \emph{EFT string limits}. Along these paths, the asymptotic scaling of the K\"ahler potential and the central charge of states charged under the $\mathrm{U}(1)$ gauge fields can be inferred from the growth theorem discussed above:

\begin{itemize}
    \item \textbf{K\"ahler potential \& distance:}\\
    Along the path \eqref{eq:path}, the asymptotic expression \eqref{Kgrowththeorem} for the K\"ahler potential reduces to
    \begin{equation}
        e^{-K}\sim \lambda^{d_{k_1\cdots k_n}}\,.
    \end{equation}
    In turn, one may readily evaluate the length $\Delta$ along the path \eqref{eq:path} as measured by the moduli space metric \eqref{eq:metric}, given by\footnote{Note that the additional factor of $\frac{1}{2}$ arises from changing to real variables. Alternatively, one may recall that the canonical normalisation for a real scalar field involves a factor of $\frac{1}{2}$. }
    \begin{equation}
    \label{eq:distance}
    \Delta = \int \sqrt{\frac{1}{2}\frac{d^2K}{d\lambda^2}} d\lambda \sim \sqrt{\frac{d_{k_1\cdots k_n}}{2}} \log\lambda\,.
    \end{equation}
    Importantly, note that the scaling coefficient in \eqref{eq:distance} is determined by the type of limiting mixed Hodge structure that appears at the \emph{endpoint} of the enhancement chain, as one would expect. Moreover, the limit $\lambda\to\infty$ is indeed at infinite distance as long as $d_{k_1\cdots k_n}\neq 0$, i.e.~the final limit is not of type I.
    \item \textbf{Central charge of particle states:}\\
    Along the path \eqref{eq:path}, the central charge \eqref{eq:BPS-mass} that vanishes fastest corresponds to a state with charge 
    \begin{equation}\label{qinGr}
        q\in \mathrm{Gr}_{3-d_{k_1}}(\Delta_{k_1})\cap\cdots \cap \mathrm{Gr}_{3-d_{k_1\cdots k_n}}(\Delta_{k_1\cdots k_n})\,,
    \end{equation}
    and scales as\footnote{Here we restrict to states for which the asymptotic coupling to the graviphoton, as defined in \cite{Bastian:2020egp}, is non-zero. As explained in \cite{Bastian:2020egp}, see also \cite{Grimm:2023lrf}, such states satisfy $Z_q\sim \mathcal{Q}_q$, such that their central charge can simply be estimated by the growth of the Hodge norm $\|q\|$ using \eqref{eq:physical-charge} and \eqref{eq:growththeorem}. While not all states satisfy this condition, we stress that those that are expected to give rise to the leading \emph{tower} do.}
    \begin{equation} \label{BPSmass-1}
        |Z_q|\sim \lambda^{-\frac{1}{2}d_{k_1\cdots k_n}}\,.
    \end{equation}
    In particular, expressing the central charge in terms of the distance via \eqref{eq:distance}, we find
    \begin{equation}
    \label{eq:BPS-mass-asymptotic}
    |Z_q|\sim e^{-\gamma \Delta}\,,\qquad \gamma = \sqrt{\frac{d_{k_1\cdots k_n}}{2}}=\begin{cases}
        \frac{1}{\sqrt{2}}\,,& d_{k_1\cdots k_n}=1\;\;\;\text{(Type II)}\,,\\
        1\,,& d_{k_1\cdots k_n}=2\;\;\;\text{(Type III)}\,,\\
        \sqrt{\frac{3}{2}}\,, & d_{k_1\cdots k_n}=3\;\;\;\text{(Type IV)}\,.
    \end{cases}
    \end{equation}
    \emph{If} for a (sub-)lattice of charges $q$ as in \eqref{qinGr} there exist BPS states, the mass of these states is given by the central charge. We can then compare these exponents with the standard expressions for the scaling of the mass of towers of states found for emergent string and decompactification limits \cite{Agmon:2022thq,Etheredge:2022opl,vandeHeisteeg:2023ubh}. Based on this, we expect the asymptotic physics of the different limits to be described by
    \begin{align}
    &\text{Type II}:\qquad \,\,\text{Emergent string limit}\,,\\
    &\text{Type III}:\qquad \text{Decompactification to 6d}\,,\\
    &\text{Type IV}:\qquad \text{Decompactification to 5d}\,,
    \end{align}
    where again we emphasise that the left-hand side refers to the \emph{endpoint} of the enhancement chain of the EFT string limits. 
    \item \textbf{EFT string tension:}\\
    Along the path \eqref{eq:path}, the tension \eqref{eq:EFT-string-tension} of an EFT string with charges $e^{k_1},\ldots, e^{k_n}$ is given by
    \begin{equation} \label{EFT-tension-limit}
    \frac{T_e}{M_{\mathrm{Pl}}^2} = -e^{k_i} \frac{\partial K}{\partial s^{k_i}}= -\frac{\partial K}{\partial\lambda} \sim \frac{d_{k_1\cdots k_n}}{\lambda}\,.
    \end{equation}
    In particular, we find
    \begin{equation}\label{scalingweight}
    \frac{m_q^2}{M_{\mathrm{Pl}}^2}\sim \left(\frac{T_e}{M_{\mathrm{Pl}}^2} \right)^{d_{k_1\cdots k_n}}\,,
    \end{equation}
    so that the integer $d_{k_1\cdots k_n}$ that characterises the endpoint of the limit is identified with the ``scaling weight'' $w$ of \cite{Lanza:2021udy}. 
\end{itemize}
We notice that for $d_{k_1\cdots k_n}=1$, the scale set by $m_q$ and $T_e$ is the same whereas for $d_{k_1\cdots k_n}>1$ the scale set by the EFT string is parametrically above the scale set by $m_q$. We recall that for a decompactification from $d\to D$ dimensions the KK-scale $m_{\rm KK}$ and the species scale $\Lambda_{\rm sp}$ (here the higher-dimensional Planck scale) are related via\footnote{The  species scale, originally introduced in~\cite{Dvali:2007hz,Dvali:2009ks,Dvali:2010vm}, has more recently been discussed in the context of the Swampland program in~\cite{Marchesano:2022axe,Castellano:2022bvr,vandeHeisteeg:2022btw,Cribiori:2022nke,vandeHeisteeg:2023ubh, Blumenhagen:2023yws,Cribiori:2023ffn,vandeHeisteeg:2023dlw,Castellano:2023aum,Castellano:2023jjt,Castellano:2023stg,Herraez:2024kux, Martucci:2024trp}.}
\begin{equation}
    \frac{m_{\rm KK}}{M_{\rm Pl}} = \left(\frac{\Lambda_{\rm sp}}{M_{\rm Pl}}\right)^{\frac{D-2}{D-d}}\,. 
\end{equation}
For decompactifications from $4\to 5$ and $4\to 6$ dimensions we thus get 
\begin{equation} \label{5d6dspeciesscales}
     \frac{m_{\rm KK,4\to 5}}{M_{\rm Pl}} = \left(\frac{\Lambda_{\rm sp}}{M_{\rm Pl}}\right)^{3} \,,\qquad  \frac{m_{\rm KK,4\to 6}}{M_{\rm Pl}} = \left(\frac{\Lambda_{\rm sp}}{M_{\rm Pl}}\right)^{2}\,,
\end{equation}
respectively. Comparing with \eqref{scalingweight}, we see that if, indeed, the limits with $d_{k_1\cdots k_n}=2$ and $d_{k_1\cdots k_n}=3$ are decompactification limits to 6d and 5d, the tension of the EFT strings realizing these limits is of the order of the respective species scale which, in these cases, is the higher-dimensional Planck scale. Thus, in the limits with $d_{k_1\cdots k_n}=2,3$ there are no light EFT strings below the relevant Planck scale. For this reason, the EFT strings in this case do not play a role for understanding the low-energy spectrum in these limits. This is a notable difference from the limits of type II with ($d_{k_1\cdots k_n}=1$), for which the EFT strings lie parametrically below the relevant Planck scale (in this case the 4d Planck scale) and represent critical emergent strings, as shown in \cite{Friedrich:2025gvs}.\newline

Let us close this section by emphasising the following important point. Purely by analysing the scaling of the couplings of the low-energy effective theory, we have found what kind of dual emergent description should arise in the various type II/III/IV limits in the moduli space. However, to actually establish this properly, one has to do more. In particular, let us point out the following two ``missing pieces''.
\begin{itemize}
    \item \textbf{Missing piece (1): Towers of states}\\
    While we have identified that the relevant light states should correspond to BPS states with charge $q\in \mathrm{Gr}_{3-d_{k_1}}(\Delta_{k_1})\cap\cdots \cap\mathrm{Gr}_{3-d_{k_1\cdots k_n}}(\Delta_{k_1\cdots k_n})$, we have not yet shown that there is in fact an infinite \emph{tower} of such BPS states. More precisely, from the EFT point of view it is not clear whether different choices of $q$ necessarily give rise to different BPS states. In fact, it is not clear if there even \emph{exists} a BPS state at all for a given choice of $q$.\footnote{As a technical remark, let us point out that since $\mathrm{Gr}_{3-d_{k_1\cdots k_n}}\cong W_{3-d_{k_1\cdots k_n}}$ and the weight filtration $W_\ell$ is defined over $\mathbb{Q}$ there are no issues of integrality.} In this regard, it is important to stress that the states corresponding to $q\in \mathrm{Gr}_{3-d_{k_1}}(\Delta_{k_1})\cap\cdots \cap\mathrm{Gr}_{3-d_{k_1\cdots k_n}}(\Delta_{k_1\cdots k_n})$ are always monodromy-invariant, so that one cannot use the arguments of \cite{Grimm:2018ohb,Grimm:2018cpv,Corvilain:2018lgw} to construct a tower using the local monodromy.\footnote{As pointed out in \cite{Grimm:2018ohb}, there is the possibility that the tower is generated by the monodromy around another divisor in the moduli space. We do not exclude this possibility. } Let us also remark that, by the results of \cite{Palti:2021ubp}, these states can only decay to states with the same mass.
    \item \textbf{Missing piece (2): Degeneracy of states}\\
    Even if one establishes the existence of a tower of states, one still has to determine that it is either a KK tower or a tower of string oscillations. To this end, given a tower of states for which the mass at level $n$ is given by $m_n$, it is important to study the \emph{degeneracy} of states $d_n$ as a function of $n$. To conclude whether a tower corresponds to a KK tower or a string tower, one has to find the behaviour
    \begin{align}
        &\text{KK tower}:\qquad d_n \sim n^{k-1}\,,\\
        &\text{string tower}:\qquad d_n\sim e^{a \sqrt{n}}\,,
    \end{align}
    for large $n$, where $k$ denotes the number of decompactifying dimensions, and $a$ is a constant whose precise value depends on the type of string. Notice that the growth of the degeneracy in the case of a KK tower is simply due to the volume of a $(k-1)$-sphere of radius $n$ such that, in the case of a KK tower, the normalized density of states 
    \begin{equation}
        \delta_n\equiv \frac{d_n}{\Omega_{k-1}}\,,
    \end{equation}
    remains constant. In particular, this means that in case we have multiple KK-$\mathrm{U}(1)$s the degeneracy of states charged under only a single of the $\mathrm{U}(1)_{\rm KK}$ factors is constant. 
\end{itemize}
The remainder of this work will be devoted to addressing these missing pieces by complementing the ``algebraic'' point of view discussed in the present section with a more geometric perspective on the type II/III/IV degenerations studied in the next section. 

\section{The II, III, IV of Calabi--Yau Degenerations}
\label{sec:II_III_IV}
In this section, we study the geometric interpretation of infinite-distance limits in the complex structure moduli space of Calabi--Yau threefolds. In the spirit of Mumford's semi-stable reduction theorem, we can, without loss of generality, focus our attention on those limits that correspond to a so-called semi-stable degeneration of the underlying family of Calabi--Yau threefolds. 
 This is because it is expected that the limits along geodesics in moduli space that we are interested in can always be embedded into a disk.
 We introduce the canonical mixed Hodge structure that is associated with the central fibre of the degeneration and explain how it relates to the limiting mixed Hodge structure introduced in Section \ref{sec:EFT-POV}. This identification is crucial in order to establish the existence of the expected KK tower(s), as will be explained in Section \ref{sec:decomp}. Additionally, we provide a simple geometric interpretation of the algebraic classification of limiting mixed Hodge structures into types II/III/IV in terms of the dimension of the so-called dual graph of the central fibre. 

\begin{figure}[t]
    \centering
    \includegraphics[width=0.65\linewidth]{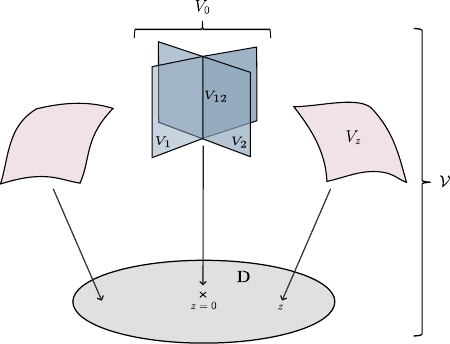}
    \caption{A schematic depiction of a semi-stable degeneration \eqref{eq:4foldV}-\eqref{eq:def-V0} with $N=2$. The fibre $V_z$ corresponding to the Calabi--Yau threefold over a generic point in the moduli space is depicted in pink, while the central fibre $V_0$ is depicted in blue. The latter splits into two components $V_1$ and $V_2$ which intersect over a complex surface $V_{12}$.}
    \label{fig:semi-stable}
\end{figure}

We start by considering a family 
\begin{equation}\begin{aligned} \label{eq:4foldV}
V_z \ \hookrightarrow & \  \ \mathcal{V} \cr 
&\ \ \downarrow\cr 
& \ \  \mathbf{D}\,
\end{aligned}\end{equation}
of Calabi--Yau threefolds varying over the unit disk $\mathbf{D}=\{z\in\mathbb{C}\,|\,|z|\leq 1\}$ with fibres $V_z$ for $z\in\mathbf{D}$. We assume that both the total space $\mathcal{V}$ as well as the generic fibres $V_z$, for $z\neq 0$, are smooth, whereas the central fibre $V_0$ is a simple normal crossing divisor of $\mathcal{V}$. The latter means that $V_0$ is a union 
\begin{equation}
\label{eq:def-V0}
    V_0 = \bigcup_{i=1}^N V_i\,
\end{equation}
of $N$ irreducible smooth components $V_i$ intersecting normally, for some positive integer $N$. Such a degeneration is referred to as a \emph{semi-stable degeneration}. By the semi-stable reduction theorem of Mumford, any degeneration for which the total space $\mathcal{V}$ is smooth and K\"ahler can be reduced to a semi-stable degeneration by performing appropriate base changes ($z\mapsto z^k$ for some integer $k$) and blowups/blowdowns of the central fibre $V_0$.\footnote{Insisting on such a strict normal crossing may be in tension with $\cV$ being Calabi--Yau. Our analysis of type III and IV limits in this work does not require $\cV$ to be Calabi--Yau. However, in~\cite{Friedrich:2025gvs} the identification of the emergent strings in type II limits made explicit use of $\cV$ being Calabi--Yau. In this case some of the components $V_i$ may be singular. We expect, however, that these singularities do not alter the analysis in~\cite{Friedrich:2025gvs}.} The situation is depicted schematically in Figure \ref{fig:semi-stable} in the case $N=2$. In addition, to compare this setup to the one introduced in Section \ref{sec:EFT-POV}, we view the fibration \eqref{eq:4foldV} as an embedding of the disk $\mathbf{D}$ into the complex structure moduli space of the family $V_z$ of Calabi--Yau threefolds. In particular, the limit $z\to 0$ corresponds to approaching some singular divisor $\Delta_{k_1\cdots k_r}$. For simplicity, we will denote the graded spaces associated with the corresponding limiting mixed Hodge structure simply by $\mathrm{Gr}_\ell H^k(V_z)$. 

Choosing local coordinates on $\mathcal{V}$ such that each divisor $V_i$ is given by $x_i=0$, a local description of the degeneration is given by 
\begin{equation}
    \prod_{i=1}^N x_i = z\,.
\end{equation}
The intersections of the various components $V_i$ will be denoted by
\begin{equation}
    V_{i_0\cdots i_k} = V_{i_0}\cap\cdots \cap V_{i_k}\,.
\end{equation}
For example, $V_{ij} = V_i\cap V_j$ denote the various double surfaces, and $V_{ijk} = V_i\cap V_j\cap V_k$ denote the various triple curves. Sometimes the spaces $V_{i_0\cdots i_k}$ are referred to as the ``codimension-$(k+1)$ strata.'' Furthermore, we introduce the notation
\begin{equation}
    V^{(k+1)} = \bigsqcup_{i_0,\ldots,i_k} V_{i_0\cdots i_k}\,,\qquad 0\leq k\leq 3\,,
\end{equation}
which collects all the strata of equal codimension into a disjoint union. 

\subsubsection*{Geometric and Algebraic Mixed Hodge Structures}
\begin{figure}
    \centering
    \includegraphics[width=\linewidth]{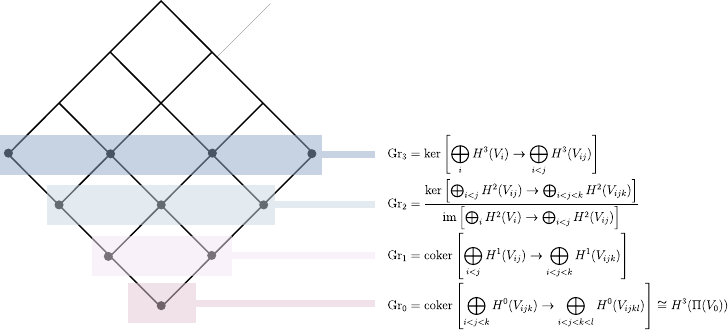}
    \caption{An overview of the various graded spaces \eqref{eq:Gr-general} that make up the geometric mixed Hodge structure on the cohomology $H^3(V_0)$ of the central fibre $V_0$. The relation $\mathrm{Gr}_0\cong H^3(\Pi(V_0))$ is elaborated upon around equation \eqref{eq:Gr0-dualgraph}.}
    \label{fig:Gr-spaces}
\end{figure}

As discussed in Section \ref{sec:EFT-POV}, we are interested in studying the cohomology of the singular variety $V_0$. Importantly, following a construction due to Deligne \cite{Deligne:1971,Deligne:1974}, each cohomology group $H^k(V_0)$, for $k=0,\ldots 3$, carries a canonical mixed Hodge structure
\begin{equation}
\label{eq:cohom-decomp}
    H^k(V_0) = \bigoplus_{\ell=0}^k \mathrm{Gr}_{\ell} H^k(V_0)\,,
\end{equation}
where the various \emph{graded spaces} $\mathrm{Gr}_\ell H^k(V_0)$, for $\ell=0,\ldots, k$, carry a pure Hodge structure of weight $\ell$. For our purposes, it will not be necessary to review this construction in detail and instead refer the reader to \cite{kulikov1998algebraic} for additional background. Put briefly, the construction employs a generalized Mayer--Vietoris spectral sequence which allows one to piece together the cohomology of $V_0$ in terms of the cohomologies of the individual components $V^{(k+1)}$ of various codimensions. The central question, then, is whether a given cohomology class restricts to a well-defined cohomology class on overlaps. The latter is characterised in terms of the so-called \v{C}ech differential $\delta$ and its associated cohomology. The various graded spaces making up the mixed Hodge structure in \eqref{eq:cohom-decomp} are then given by
\begin{equation}
\label{eq:Gr-general}
    \mathrm{Gr}_\ell H^{k}(V_0) = \frac{\mathrm{ker}\left[\delta:H^\ell(V^{(k-\ell+1)})\rightarrow H^\ell(V^{(k-\ell+2)})\right]}{\mathrm{im}\left[\delta:H^\ell(V^{(k-\ell)})\rightarrow H^\ell(V^{(k-\ell+1)})\right]} \,,\qquad \ell=0,\ldots, k\,.
\end{equation}
In words, the space $\mathrm{Gr}_\ell H^{k}(V_0)$ consists of $\ell$-cocycles on the codimension-$(k-\ell)$ components $V^{(k-\ell+1)}$ that do not arise from the restriction of $\ell$-cocycles in one dimension higher, but do restrict to $\ell$-cocycles in one dimension lower. In Figure \ref{fig:Gr-spaces} we give an overview of the various expressions for the graded spaces in the case $k=3$.

\paragraph{Example:} Let us exemplify the formula \eqref{eq:Gr-general} for $k=3$, in the case where all triple intersections $V_{ijk}$ are empty, such that $V^{(q)}=\emptyset$ for $q\geq 2$. One finds
\begin{align}
    \mathrm{Gr}_3 H^3(V_0) &= \mathrm{ker}\left[\delta: \bigoplus_{i} H^3(V_{i})\to\bigoplus_{i<j} H^3(V_{ij})  \right]\,,\\
    \mathrm{Gr}_2 H^3(V_0) &= \mathrm{coker}\left[\delta: \bigoplus_{i} H^2(V_{i})\to\bigoplus_{i<j} H^2(V_{ij})  \right]\,,\\
    \mathrm{Gr}_1 H^{3}(V_0) &=\emptyset\,,\\
    \mathrm{Gr}_0 H^{3}(V_0) &=\emptyset\,.
\end{align}
In particular, the space $\mathrm{Gr}_3 H^3(V_0)$ consists of 3-cocycles on the components $V_i$ that restrict to 3-cocycles on the double surfaces $V_{ij}$. Similarly, the space $\mathrm{Gr}_2 H^3(V_0)$ consists of 2-cocycles on the double surfaces $V_{ij}$ that do not arise as the restriction of 2-cocyles on the components $V_i$. 

In general, the geometric mixed Hodge structure described above is not identical to the limiting mixed Hodge structure described in Section \ref{sec:EFT-POV}. The relation between the two is made precise by the so-called Clemens--Schmid long exact sequence. Again, it will not be necessary to describe this in full detail, and we instead refer the interested reader to \cite{Morrison:1984}. For our purposes, the central result is captured by Corollary 1 in \textit{loc.~cit.}, which may be summarized as
\begin{align}
\label{eq:weight-geometric-limiting}
    W_\ell H^3(V_0) \cong W_\ell H^3(V_z) \cap \mathrm{ker}\,N\,,\qquad 0\leq \ell<3\,,
\end{align}
where we recall that $W_\ell H^3(V_0)$ and $W_\ell H^3(V_z)$ denote the weight filtrations associated to the geometric mixed Hodge structure and the limiting mixed Hodge structure, respectively. Furthermore $N$ denotes the log-monodromy operator.\footnote{Recall that the graded spaces are constructed from the weight filtration via $\mathrm{Gr}_\ell = W_\ell / W_{\ell-1}$.} In particular, the geometric mixed Hodge structure only captures the monodromy-invariant part of the limiting mixed Hodge structure. However, as will become clear in Section \ref{sec:decomp}, the \emph{leading} tower of states is always monodromy-invariant. In particular, for the purpose of understanding the emergent dual frame in a given degeneration, we may simply identify the geometric and limiting mixed Hodge structures, at least for $\ell<3$. As a final comment, let us stress that the relation \eqref{eq:weight-geometric-limiting} becomes more involved for $\ell=3$ (see Corollary 3 in \textit{loc.~cit.}), but this will not be relevant for our discussion as such states will never be part of the leading tower.  

\subsubsection*{Geometric Interpretation of II/III/IV}
Having established the relation \eqref{eq:weight-geometric-limiting} between the geometric and limiting mixed Hodge structures, we can now employ \eqref{eq:Gr-general} to study the geometric interpretation of the algebraic classification into types II, III, and IV, introduced in Section \ref{sec:EFT-POV}. To this end, we make the following two observations.
\begin{itemize}
    \item \textbf{Observation (1):}\\
    If $V^{(q+1)}=\emptyset$, then it follows from equation \eqref{eq:Gr-general} that $\mathrm{Gr}_{3-q} H^{3} (V_0)=\emptyset$. In other words, in order for the degeneration to be of type II/III/IV, there must exist a non-empty double surface $V_{ij}$ / triple curve $V_{ijk}$ / quadruple point $V_{ijkl}$, respectively. 
    \item \textbf{Observation (2):}\\
    The space $\mathrm{Gr}_0 H^k(V_0)$ can alternatively be described in terms of the cohomology of the so-called \emph{dual graph} of the normal crossing divisor $V_0$. The latter is a simplicial complex, denoted by $\Pi(V_0)$, that is constructed as follows. First, to each component $V_i$ one assigns a vertex. Then, for each connected component in the intersection $V_{ij}=V_i\cap V_j$ one assigns an edge connecting the respective vertices. Continuing on, one inductively assigns a $k$-simplex to each connected component of a non-empty intersection $V_{i_0\cdots i_k}$. One can then show that, in terms of the dual graph, the space $\mathrm{Gr}_0 H^k(V_0)$ is given by
    \begin{equation}
    \label{eq:Gr0-dualgraph}
        \mathrm{Gr}_0 H^k(V_0) = H^k(\Pi(V_0))\,,
    \end{equation}
    see for example \cite{Morrison:1984,kulikov1998algebraic} for further details. Notably, since we are interested in complex structure degenerations of the Calabi--Yau threefold only the space $H^3(V_0)$ acquires a non-trivial mixed Hodge structure, such that by \eqref{eq:Gr0-dualgraph} the Betti numbers of $\Pi(V_0)$ read
    \begin{equation}
        (b_0,b_1,b_2) = (1,0,0)\,.
    \end{equation}
    Additionally, if the degeneration is of type II or III then by \eqref{eq:Gr0-dualgraph} also $b_3=0$, while if the degeneration is of type IV then $b_3=1$. Some examples of dual graphs arising in two-parameter families of Calabi--Yau threefolds will be discussed momentarily. 
\end{itemize}
The above observations can be summarized into the following \emph{geometric} characterisation of the classification of limiting mixed Hodge structures:
\begin{equation}
    \boxed{\text{Type II / III / IV}\qquad \implies\qquad \text{$\Pi(V_0)$ has dimension 1/2/3\,.}}
\end{equation}
In other words, whether the degeneration is of type II/III/IV is directly reflected in the intersection pattern of the components appearing in the semi-stable degeneration. In addition, the topology of $\Pi(V_0)$ is restricted as explained in Observation (2) above. Let us briefly describe and exemplify this geometric characterisation for each of the three types of mixed Hodge structures. 
\begin{itemize}
    \item \textbf{Type II:}\\
    The space  ${\rm Gr}_2 \neq \emptyset$, but ${\rm Gr}_1 = \emptyset = {\rm Gr}_0$.
    The dual graph is a one-dimensional simplicial complex with $(b_0,b_1)=(1,0)$. In other words, $\Pi(V_0)$ is a connected acyclic graph, i.e.~a \emph{tree graph}. 
    \paragraph{Example:}
    For an example of a type II degeneration, consider the mirror of the Calabi--Yau $\mathbb{P}^{1,1,2,2,6}[12]$ defined as a degree-12 hypersurface in weighted projective space given by the equation
    \begin{equation}
    \label{eq:II-example}
        P = x_1^{12}+x_2^{12}+x_3^6+x_4^6+x_5^2 -12\psi\, x_1 x_2 x_3 x_4 x_5-2\phi\, x_1^6 x_2^6=0\,.
    \end{equation}
    To be precise, the mirror is obtained by taking the quotient $\{P=0\}/(\mathbb{Z}_6^2\times \mathbb{Z}_2)$ and resolving all orbifold singularities \cite{Candelas:1993dm}. In the limit $\phi\to\infty$ the Calabi--Yau degenerates into two irreducible components
    \begin{equation}
        V_0 = V_1\cup V_2\,,\qquad V_i = \{x_i^6=0\}\,,\qquad i=1,2\,.
    \end{equation}
    Additionally, the double surface $V_{12}=V_1\cap V_2$ is non-empty and is in fact a K3-surface. Hence, this is indeed an example of a type II limit, and the dual graph is simply an interval. 
    \item \textbf{Type III:}\\
    The space 
   ${\rm Gr}_1 \neq \emptyset$,
   but ${\rm Gr}_0 = \emptyset$.
    The dual graph is a two-dimensional simplicial complex with $(b_0,b_1,b_2)=(1,0,0)$. For example, $\Pi(V_0)$ could be a triangulation of a two-dimensional disk. 
    \paragraph{Example:}
    For an example of a type III degeneration, consider the mirror of the Calabi--Yau $\mathbb{P}^{1,1,1,6,9}[18]$ given by a degree-18 hypersurface in weighted projective space defined by the equation
    \begin{equation}
    \label{eq:III-example}
        P = x_1^{18}+x_2^{18}+x_3^{18}+x_4^3+x_5^2 -18\psi\, x_1 x_2 x_3 x_4 x_5-3\phi\, x_1^6 x_2^6 x_3^6=0\,.
    \end{equation}
    Again, to be precise, the mirror is obtained by taking the quotient $\{P=0\}/(\mathbb{Z}_{18}\times \mathbb{Z}_6)$ and resolving all orbifold singularities \cite{Candelas:1994hw}. In the limit $\phi\to\infty$ the Calabi--Yau degenerates into three irreducible components
    \begin{equation}
        V_0 = V_1\cup V_2\cup V_3\,,\qquad V_i = \{x_i^6=0\}\,,\qquad i=1,2,3\,.
    \end{equation}
    Additionally, the triple curve $V_{123}=V_1\cap V_2\cap V_3$ is non-empty and is in fact a two-torus $T^2$. Hence, this is indeed an example of a type III limit, and the dual graph is a solid triangle. 
    \item \textbf{Type IV:}\\
  The space
   ${\rm Gr}_0 \neq \emptyset$,
   but ${\rm Gr}_1 = \emptyset$.
    The dual graph is a two-dimensional simplicial complex with $(b_0,b_1,b_2,b_3)=(1,0,0,1)$. In other words, $\Pi(V_0)$ is a $\mathbb{Q}$-homology 3-sphere (i.e.~a 3-manifold with the same rational homology as a 3-sphere). In fact, it has been shown that, under reasonable assumptions\footnote{To be precise, the central fibre is assumed to be a union of logCY pairs.}, the dual graph $\Pi(V_0)$ is homeomorphic to a standard 3-sphere, i.e. \cite{Kollar:2015}
\begin{equation}
    \Pi(V_0)\cong S^3\,.
\end{equation}

    \paragraph{Example:}
    Both the examples given in \eqref{eq:II-example} and \eqref{eq:III-example} give rise to a type IV degeneration in the limit $\psi\to\infty$, where the Calabi--Yau degenerates into five irreducible components
    \begin{equation}
        V_0 = V_1\cup \cdots\cup V_5\,,\qquad V_i = \{x_i=0\}\,,\qquad i=1,\ldots 5\,.
    \end{equation}
    Additionally, all five quadruple points $V_{ijkl}$ are non-empty. Hence, this is indeed an example of a type IV limit, and the dual graph can be shown to be the standard 3-simplex, which is topologically equivalent to the standard 3-sphere $S^3$. 
\end{itemize}
As a final remark, let us note that the role of the dual graph may be slightly surprising from an algebro-geometric point of view. To clarify this, one should instead adopt a differentio-geometric point of view, taking into account the behaviour of the \emph{metric} on the Calabi--Yau $V_z$ as we approach $z\to 0$. This point of view is crucial in intrinsically geometric approaches to studying mirror symmetry following the works of Gross--Siebert initiated in \cite{GrossSiebert2003}. Although it is not required for the rest of this work, we refer the interested reader to Appendix \ref{app:dual-graph} for additional background on this.

\section{Type III and IV as Decompactification Limits}
\label{sec:decomp}

In this section, we combine the perspectives introduced in Sections \ref{sec:EFT-POV} and \ref{sec:II_III_IV} to show that limits of type III/IV are indeed decompactification limits to 6d/5d. The general strategy is the same for both types of limits and follows the reasoning first used in \cite{Friedrich:2025gvs} to address the interpretation of type II limits as emergent string limits. The central observation is the following. Geometrically, the (putative) BPS particles discussed in Section \ref{sec:EFT-POV} arise from D3-branes wrapping a special Lagrangian 3-cycle $\Gamma$ that is Poincar\'e dual to a charge $q\in H^3(V,\mathbb{Z})$. When this charge additionally lies in the space $\mathrm{Gr}_{3-d}$, the volume of $\Gamma$ vanishes asymptotically as
\begin{equation}
    \mathrm{Vol}(\Gamma) = \frac{m_q}{M_{\rm Pl}}\sim\lambda^{-d/2}\,,
\end{equation}
as follows from (\ref{eq:growththeorem}). 
At the same time, from the discussion in Section \ref{sec:II_III_IV}, we know that the space $\mathrm{Gr}_{3-d}$ roughly consists of certain $(3-d)$-cycles that remain in the central fibre $V_0$.\footnote{For notational simplicity we drop the additional labels distinguishing whether $\mathrm{Gr}_{3-d}$ refers to the geometric or limiting mixed Hodge structure, since we have argued in Section \ref{sec:II_III_IV} that, for our purposes, the two may be identified.} Importantly, one can think of these as coming from 3-cycles in $V_z$ for which $d$ ``legs'' shrink as one approaches the infinite-distance limit, with each leg shrinking at a rate $\lambda^{-1/2}$ as $\lambda\to\infty$. By using the properties of the geometric mixed Hodge structure and the triviality of certain normal bundles, we argue that the topology of the 3-cycle $\Gamma$ can locally be thought of as a torus-fibration
\begin{equation}
    T^{d}\hookrightarrow \Gamma\to B_{3-d}\,,
\end{equation}
over a real $(3-d)$-dimensional base $B_{3-d}$, with the $T^d$-fibre shrinking to zero size at the degeneration. More precisely, for type III limits $(d=2)$ we establish that there are precisely two such fibrations giving rise to the two KK towers expected for decompactification limits to 6d, while for type IV limits $(d=3)$ there is a unique such fibration signaling a decompactification limit to 5d. We discuss these two cases in turn below. Additionally, we argue that in both limits the BPS indices $\Omega_{\rm BPS}(n\Gamma)$ are given by
\begin{equation}
    \Omega_{\rm BPS}(n\Gamma) = -\chi(V)\,,\quad \forall n\in\mathbb{Z}_{>0}\,,
\end{equation}
where the right-hand side is given by (minus) the Euler character of the Calabi--Yau threefold $V$. Here, following \cite{Banerjee:2022oed}, the BPS index for A-branes wrapping a special Lagrangian 3-cycle $\Gamma$ is defined as 
\begin{equation}\label{def:OmegaBPS}
    \Omega_{\rm BPS}(\Gamma) = (-1)^{\cM_\Gamma} \chi(\cM_{\Gamma})\,,
\end{equation}
with $\cM_\Gamma$ the moduli space of the A-brane on $\Gamma$, and $\chi(\cM_\Gamma)$ its Euler characteristic. 

Our geometric results then provide the two missing pieces discussed in Section~\ref{sec:EFT-POV}. First, they establish the existence of a tower of BPS states in the relevant graded spaces, and second, they show that the degeneracy density is constant as expected for KK states charged under a single KK-$\rm U(1)$.

\subsection{Type III}
\label{subsec:III}

Consider a type III limit.
 As discussed in the previous two sections, the smallest value of $d$ such that the graded space ${\rm Gr}_{3-d} \neq \emptyset$ is $d =2$, with ${\rm dim}\,{\rm Gr}_1 = 2$. This space contains the candidate charges of the leading tower.
Fix a seed charge $q_0\in H^3(V,\mathbb{Z}) \cap \mathrm{Gr}_1$. Using the general result \eqref{eq:Gr-general}, we have
\begin{equation}
\label{eq:Gr1}
    \mathrm{Gr}_1\cong \frac{\bigoplus_{i<j<k} H^1(V_{ijk})}{\mathrm{im}\left[\delta:\bigoplus_{i<j}H^1(V_{ij})\to \bigoplus_{i<j<k}H^1(V_{ijk}) \right]}\,.
\end{equation}
In words, this means that $q_0$ can be viewed as a harmonic 1-form on the triple curves $V_{ijk}$ that does not arise from the restriction of a harmonic 1-form on the double surfaces $V_{ij}$. At first sight, this space may appear very complicated. However, it is severely restricted due to the Calabi--Yau condition, which enforces that $\mathrm{dim}\,\mathrm{Gr}_1 = 2$. This implies that at least for one of the triple curves $V_{ijk}$, its first cohomology group must contain the first cohomology of a smooth elliptic curve.
  In fact, when the double surfaces are rational, these triple curves must be elliptic because $h^{1,0}(V_{ij}) = 0$, as is the case 
   in the example \eqref{eq:III-example}.\footnote{In addition, this typically happens in a Tyurin degeneration in which the K3 additionally undergoes a type II Kulikov degeneration. Another possibility is that there are multiple $T^2$ factors appearing among the triple curves, as might happen when the double surfaces are elliptic ruled surfaces, but note that all of these either have to have the same complex structure or are quotiented out as in \eqref{eq:Gr1} in order to have $\mathrm{dim}\,\mathrm{Gr}_1=2$. } 
   For simplicity of presentation we will in the sequel assume this, but our conclusions do not rely on this assumption.
   Finally, we cannot exclude that some of the triple curves have the topology of a $\mathbb{P}^1$ since, being topologically trivial, they would not contribute to the mixed Hodge structure. The upshot of this discussion is that effectively
\begin{equation}
    \mathrm{Gr}_1 \cong H^1(T^2)\,.
\end{equation}
Let us denote the generators of $H^1(T^2)$ by $\omega_A$ and $\omega_B$, with dual 1-cycles $\gamma^A_1$ and $\gamma_1^B$. 

We can now construct the 3-cycle $\Gamma_0$ that is Poincar\'e dual to $q_0$. To this end, we first recall that $V_0$ is a normal intersection variety inside a smooth Calabi--Yau fourfold $V_0 \hookrightarrow \cV \to \mathbf{D}$ which can be written as
\begin{equation}
    V_0 = \{z=0\} \subset \cV\,,
\end{equation}
where we recall that $z$ is the coordinate on the unit disk $\mathbf{D}$. Similarly, due to the normal crossing property, we can locally write 
\begin{equation}
    V_i = \{x_i=0\} \subset V_0\,,
\end{equation}
where $x_i$ is a local coordinate on $V_0$. Inside $V_i$, the triple curve $V_{ijk}$ is globally cut out as the locus
\begin{equation}
    V_{ijk} = \{x_j=0\} \cap \{x_k=0\} \subset V_i\,. 
\end{equation}
For this reason, the normal bundle of $V_{ijk}$ inside $V_i$ is trivial 
\begin{equation}
    N_{V_{ijk}|V_i} = \mathcal{O}_{V_{ijk}}^{\oplus 2}\,.
\end{equation}
Repeating this for $V_j$ and $V_k$, we find that the normal bundles in each $V_{ijk}\subset V_{j}$ and $V_{ijk}\subset V_k$ are also trivial. The normal bundle of $V_{ijk}$ is thus trivial in all components of $V_0$ that contain $V_{ijk}$. We can therefore conclude that the normal bundle of $V_{ijk}$ inside $V_0$ is also trivial
\begin{equation}
    N_{V_{ijk}|V_0}= \mathcal{O}_{V_{ijk}}^{\oplus 2}\,.
\end{equation}
Locally in the vicinity of $V_{ijk}$, the geometry of $V_0$ can thus be viewed as the total space of the bundle
\begin{equation}\label{bundleTypeIII}
    \mathcal{O}^{\oplus 2}_{V_{ijk}} \to V_{ijk} \,.
\end{equation}
Since the bundle is trivial, it is in fact a direct product of two copies of $\mathbb{C}^*$ and $V_{ijk}$. 

The topology of the 3-cycles shrinking at a type III degeneration can be inferred from this local geometry. To this end, consider the circle $S^1=\{z=re^{i \varphi} \in \mathbb{C}^*|\varphi\in [0,2\pi)\}$ in either of the copies of $\mathbb{C}^*$ in the total space of the bundle in \eqref{bundleTypeIII}. Taking such a circle, we can fiber it over a 1-cycle $\gamma_1\in H_1(V_{ijk})$. Since the fibration in \eqref{bundleTypeIII} is trivial, the resulting 2-cycle is topologically $S^1\times S^1$. Together with a similar 1-cycle in the second $\mathbb{C}^*$-factor in \eqref{bundleTypeIII} we then get a 3-cycle that is topologically a $T^3$. The volume minimizing 3-cycle then corresponds to the limit $r\to 0$ where the $S^1$ in the $\mathbb{C}^*$-fibres is contracted to the origin. Thus, the volume minimizing 3-cycle is a 3-cycle that shrinks to the 1-cycle $\gamma_1$. 

Since the triple curve $V_{ijk}$ is a torus, $b_1(V_{ijk})=2$, such that there is a two-dimensional lattice of 1-cycles $\gamma_1 \in H_1(V_{ijk})$. By our above argument we thus get a two-dimensional lattice of shrinking $T^3$s. Let $\gamma_1^A$ and $\gamma_1^B$ be the primitive generators of $H_1(V_{ijk})$. These give rise to the generators $\Gamma_0^{(A)}$ and $\Gamma_0^{(B)}$ of the lattice of shrinking $T^3$s obtained as linear combinations 
\begin{equation}
    \Gamma_0^{(m,n)} = m \Gamma_0^{(A)} + n \Gamma_0^{(B)}\,. 
\end{equation}
The volume of each of these 3-cycles vanishes at the same rate in $\lambda$ but their volume with respect to $\Gamma_0^{(A)}$ is determined by the complex structure $\tau$ of $V_{ijk}$: %\textcolor{orange}{JM: should probably be $\mathrm{Im}\,\tau$ instead of $|\tau|$?}
\begin{equation}\label{eq:relativevolume}
    \frac{\cV_{\Gamma_0^{(m,n)}}}{\cV_{\Gamma_0^{(A)}}} = |m+n\tau|\,.
\end{equation}
Since all $\Gamma_0^{(m,n)}$ have the topology of a $T^3$, the corresponding BPS invariants of wrapped D3-branes are constant, i.e.
\begin{equation}
    \Omega_{\rm BPS}(k\Gamma_0^{(m,n)}) = \Omega_{\rm BPS}(\Gamma_0^{(m,n)})\,,\qquad \forall k\in \mathbb{N}_{>0}\,. 
\end{equation}
Notice that, unlike in the type II case, there are no other 3-cycles with $b_1(\Gamma_0)\geq 3$ that shrink at the same rate as $\Gamma_0^{(m,n)}$ such that, in this case, there is indeed just a two-dimensional tower of light states arising at the type III singularity for which the BPS invariants do not grow with increasing wrapping number. 

A priori, the BPS invariants for $\Gamma_0^{(A)}$ can differ from $\Gamma_0^{(B)}$ such that, more generally, the BPS invariants of $\Gamma_0^{(m,n)}$ depend on the choice of $(m,n)$. However, as a result of the analysis in the next section, we will see that consistency of the underlying theory of quantum gravity requires $\Omega_{\rm BPS}(\Gamma_0^{(A)})=\Omega_{\rm BPS}(\Gamma_0^{(B)})$ implying that also $\Omega_{\rm BPS}(\Gamma_0^{(m,n)})$ is independent of $(m,n)$. 

The basic insight allowing us to relate the BPS invariants to each other is that every type III singularity can be enhanced to a type IV singularity in two different ways.\footnote{This was already speculated from a different point of view in \cite{Bastian:2020egp}.} To see this, recall that the triple-curve $V_{ijk}$ arising at a type III singularity is a genus-one curve. The complex structure $\tau$ of this curve can be identified with one of the residual moduli along the type III singularity. The extreme limits $\tau\to i\infty$ and $\tau\to 0$ correspond to further degenerations of the torus. This additional degeneration enhances the type III limit to type IV limits since (after stable degeneration) the torus splits into at least two components intersecting over points. From \eqref{eq:relativevolume} we infer that one of the two $T^3$s arising at the type III singularity has a parametrically smaller volume. Without loss of generality, we take this to be $\Gamma_0^{(A)}$ which is then the \emph{unique} 3-cycle with $b_1(L)\geq 3$ that becomes light at the limit realizing a type III $\to$ IV enhancement. Based on a physical consistency argument, we argue in the next section that, for a given Calabi--Yau threefold, the BPS invariants of D3-branes wrapping the $T^3$ shrinking at \emph{any} type IV limit in the complex structure moduli space of this threefold have to be the same. 

\subsection{Type IV}
\label{subsec:IV}

The local geometry of type IV singularities has been discussed in Section~\ref{sec:II_III_IV} with $\text{Gr}_0H^3 (V_0)$ being non-zero and corresponding to the third cohomology of a three-sphere. As for the type III degeneration, we can zoom in to the local geometry in the vicinity of the quadruple intersection points $V_{ijkl}$. Since these are just points, the local geometry can be viewed as the total space of the normal bundle $\mathcal{O}(0)^{\oplus 3}$, or as three copies of $\mathbb{C}^*$. As a consequence, we now have a single $T^3$ that shrinks over the quadruple point from the three $S^1$s in $\mathbb{C}^*$ around the origin. This $T^3$ is the 3-cycle that vanishes at the fastest rate in the type IV degeneration limit. Let us denote this $T^3$ by $\Gamma_0$. As before, 
\begin{equation}
    \Omega_{\rm BPS}(n\Gamma_0) = \Omega_{\rm BPS}(\Gamma_0)\,,
\end{equation}
which follows from $\Gamma_0$ being a $T^3$. Thus, the BPS invariants for the ray in the charge lattice along the $\Gamma_0$ directions are constant. Physically, we can interpret these states as KK modes of a decompactification to five dimensions where the 1-cycle that is becoming large has an isometry leading to the KK-$\rm U(1)$ under which the tower of states is charged. 

By~\eqref{def:OmegaBPS}, the invariant $\Omega_{\rm BPS}$ is closely related to the Euler characteristic of the moduli space of the 3-cycle $\Gamma_0$ as a special Lagrangian submanifold. The moduli space of $\Gamma_0$ depends on the details of the embedding of the degenerate $T^3$ inside the full $V_0$. This embedding is different for each type IV limit that we can consider, such that, a priori, also the Euler characteristic of the moduli space of $\Gamma_0$ may differ; this would lead to different values of $\Omega_{\rm BPS}(\Gamma_0)$ depending on which type IV singularity is considered. In particular, for a given threefold $V$ there can be multiple, disjoint type IV divisors in its complex structure moduli space for which a different $T^3$ vanishes. The Euler characteristic of the moduli space of these different $T^3$s inside a given threefold $V$ can then in principle differ. 

To study how the moduli spaces of the associated special Lagrangian 3-cycles differ from a geometric perspective would require additional information about the details of the degeneration. Instead, here we use a physical argument to determine $\Omega_{\rm BPS}(\Gamma_0)$ for \emph{any} type IV singularity. To achieve this, the key physical insight is that the D3-branes multi-wrapping $\Gamma_0$ correspond to the KK tower of massless, uncharged fields in a 5d parent theory. The BPS index of these KK-modes is then related to the BPS index of the massless, uncharged states in the 5d theory.

In five dimensions, the BPS index can be computed from the helicity supertrace, $B_2$, over the representations of the $SU(2)_R$ symmetry. In five dimensions the Lorentz group decomposes as $SO(4)\cong SU(2)_L \times SU(2)_R$ such that the supertrace of the massless, uncharged fields in 5d can be written as
\begin{equation}
    (B_2)_{\rm BPS, 5d}^{(0)} = \sum_{j_R} (-1)^{2j_R}(2j_R+1) n_{(j_L,j_R)}^{0}\,,
\end{equation}
where $n_{(j_L,j_R)}^{0}$ are the massless, uncharged representations of the 5d parent theory emerging as we approach the type IV divisor. To obtain from this the BPS index defined in~\eqref{def:OmegaBPS} via the Witten index of the SQM, we have to factor out the contribution of the center of mass to the supertrace (see e.g. \cite{Alexandrov:2020qpb}). Since the states we are interested in are $\frac12$-BPS, factoring out the center of mass contribution gives an overall sign such that
\begin{equation}
   \Omega_{\rm BPS, 5d}^{(0)} = - \sum_{j_R} (-1)^{2j_R}(2j_R+1) n_{(j_L,j_R)}^{0}\,.
\end{equation}
By construction, we know that upon compactification on some one-dimensional manifold, the five-dimensional theory gives rise to a 4d $\cN=2$ theory. Therefore, the 5d theory itself has to preserve at least eight supercharges. For this reason, we can use the 5d $\cN=1$ multiplets to characterize the $SU(2)_R$ representations. The fermions in the respective 5d $\cN=1$ multiplet contribute as 
\begin{equation}
    \quad (j_L,j_R)=\left\{\begin{matrix} (0,\frac32)&  \text{supergravity}\,,\vspace{6pt} \\  (0,\frac12)& \text{vector} \,,\vspace{6pt}\\ (0,0)\times 2&\text{hyper} \,,\end{matrix}\right.
\end{equation}
such that for the 5d theory emerging in a type IV limit with $n_{V}^{(0)}$ massless vector multiplets and $n_{H}^{(0)}$ massless, neutral hypermultiplets we get 
\begin{equation}
   \Omega_{\rm BPS, 5d}^{(0)} = -2 n_{H}^{(0)} + 2 n_{V}^{(0)} +4 \,. 
\end{equation}
To relate $\Omega_{\rm BPS, 5d}^{(0)}$ with the BPS index of the KK-modes of the massless, uncharged 5d fields, we first have to specify the compactification manifold. The compactification of the 5d theory has to satisfy two properties: 
\begin{enumerate}
    \item It preserves eight supercharges. 
    \item There is a $\rm U(1)$ gauge symmetry that acts like a KK-$\rm U(1)$. 
\end{enumerate}
These two requirements uniquely fix the compactification space to be a circle since orbifolds would break the isometry, removing the KK-$\rm U(1)$ that is supposed to act as the central charge of the KK-modes of \emph{uncharged} fields in five dimensions. \emph{If} all massless 5d fields give rise to massless fields after decompactification, we can relate $\Omega_{\rm BPS,5d}^{(0)}$ to the Hodge numbers of the Calabi--Yau threefold using 
\begin{equation}
    n_{H}^{(0)} = h^{1,1}(V) +1\,, \quad \text{and} \quad n_V^{(0)} = h^{2,1}(V) -1 \,,
\end{equation}
and it follows that 
\begin{equation}
    \Omega_{\rm BPS, 5d}^{(0)}= -2\left(h^{1,1}(V)-h^{2,1}(V) \right) = -\chi(V)\,,
\end{equation}
where $\chi(V)$ denotes the Euler characteristic of $V$. If in addition each massless, uncharged field also gives rise to a BPS KK tower, the BPS index at each level within this tower is given by $\Omega_{\rm BPS, 5d}^{(0)}$ and hence by the Euler characteristic of $V$. 
 The two conditions for this to happen are therefore: 
\begin{itemize}
    \item All massless, uncharged fields in the five-dimensional parent theory have zero modes along the $S^1$. 
    \item All massless, uncharged fields give rise to a BPS KK tower. 
\end{itemize}
Let us first address the existence of a zero mode. For a massless field in the five-dimensional theory to not give rise to a zero mode upon circle compactification, we have to give a vev to one of the massless fields in the five-dimensional theory, effectively generating a mass term in the 4d action. The only fields that can be given a vev are the scalars and the vectors (i.e.~a Wilson line) in the 5d multiplets. Since the 5d $\cN=1$ theory does not have a non-trivial scalar potential, only derivatives of the scalar fields appear. Thus, to get a non-trivial mass term for any of the zero modes of the 5d fields, the scalar fields have to have a non-trivial profile along the $S^1$. Such a profile would, however, break the isometry along the $S^1$ and hence the KK-$\rm U(1)$. 

Thus, the only possibility to lift zero modes is via a non-trivial Wilson line for one of the gauge fields in the vector multiplets of the 5d theory. For a field $\Phi(x^\mu, y)$ with charge $q$ under a five-dimensional $\rm U(1)$ gauge field $A_M$ for which we turn on a Wilson line $A_y$, the four-dimensional equations of motion read 
\begin{equation}
    \Delta_4 \Phi_n(x^\mu) = \left(\frac{n}{R}-qA_y\right)^2 \Phi_n(x^\mu)\,,
\end{equation}
where $\Phi_n$ denotes the $n$-th KK mode, $R$ denotes the radius of $S^1$. In this case, the zero mode ($n=0$) gets a non-zero mass $m^2_0 = (qA_y)^2$. However, this only works for fields that are \emph{charged} in the five-dimensional parent theory. To determine the BPS index  of $\Gamma_0$ instead we are interested only in the uncharged fields. 

Since for uncharged fields in the 5d $\cN=1$ parent theory, there is no way to introduce an effective mass of the form 
\begin{equation}
    \Delta_4 \Phi_n(x^\mu) = \left(\frac{n}{R}+\tilde m\right)^2 \Phi_n(x^\mu)\,,
\end{equation}
without breaking supersymmetry of the effective 4d theory and/or breaking the KK-$\rm U(1)$, also the mass of the KK modes is uncorrected,
\begin{equation}
    M_n^2 = \frac{n^2}{R^2}\,. 
\end{equation}
The RHS agrees with the physical KK charge of the state, such that the KK-$U(1)$ acts as a central charge and, accordingly, the KK modes are BPS. 

We thus conclude that for each massless, uncharged field in the 5d $\cN=1$ parent theory, supersymmetry and the requirement to have a KK-$\rm U(1)$ ensures that upon $S^1$ compactification we obtain a zero mode and a tower of BPS KK-modes. Thus, the BPS index of the zero modes is the same as the BPS index at each level of the KK tower and given by $\Omega_{\rm BPS, 5d}^{(0)}$. For the 3-cycle $\Gamma_0$ arising at a type IV singularity this thus means 
\begin{equation}
    \Omega_{\rm BPS}(n\Gamma_0) = -\chi(V) \,,\quad \forall n\in \mathbb{Z}_{> 0}\,. 
\end{equation}
For the type IV singularity arising at the large complex structure point of $V$ this result is well known~\cite{Strominger:1996it}: mirror symmetry maps the D3-brane on $\Gamma_0$ to the D0-brane of Type IIA on the mirror $\hat{V}$. The moduli space of the D0-brane is thus $\hat{V}$ such that
\begin{equation}
    \Omega_{\rm BPS}({\rm pt}) =  \chi(\cM_{\rm pt}) = \chi(\hat{V}) =-\chi(V)\,. 
\end{equation}
We stress that our result is stronger since it $i)$ does not rely on the realisation of mirror symmetry as a triple T-duality and $ii)$ also applies to type IV singularities that do not intersect the large complex structure divisor in moduli space where mirror symmetry is hence not available.

As mentioned at the end of Section~\ref{subsec:III}, any type III singularity can be enhanced to a type IV singularity by degenerating the torus arising as a triple curve of a type III degeneration. This identifies the BPS degeneracy of the two $T^3$s, $\Omega_{\rm BPS}(\Gamma_0^{(A)})$ and $\Omega_{\rm BPS}(\Gamma_0^{(B)})$, with the degeneracy of the single $T^3$ arising at a type IV singularity. By our above reasoning, also for type III degenerations the BPS degeneracy of (multi-wrappings of) the two $T^3$s is given by the Euler characteristic of $V$. This is summarised as the following statement about the BPS indices for vanishing 3-tori:  \\

\begin{mdframed}[backgroundcolor=white,shadow=true,shadowsize=4pt,shadowcolor=black,roundcorner=6pt]
When the complex structure of a Calabi-Yau threefold $V$ undergoes a degeneration $\Delta_{k_1 \ldots k_n}$ of type III or of type IV, the BPS index of each of the two shrinking 3-tori $\Gamma_0^{A}$, $\Gamma_0^{B}$ in ${\rm Gr}_1(\Delta_{k_1 \ldots k_n})$ (type III) or of the single shrinking 3-torus $\Gamma_0$ in ${\rm Gr}_0(\Delta_{k_1 \ldots k_n})$ (type IV), respectively, is given by 
\begin{equation}
\Omega_{\rm BPS}(m \Gamma_0) = \Omega_{\rm BPS}(m \Gamma_0^{(i)}) =- \chi(V)    \,,  \quad \forall\,   m \in \mathbb Z_{m>0} \,, \qquad i = A,B     \,.
\end{equation}

\end{mdframed}

\section{Geometry and Physics of Singularity Enhancements} \label{sec_Enhancements}
Up to now, we have focused on paths in the moduli space approaching singular divisors along EFT string limits, for which all moduli that approach an extreme limit are scaled at the same rate, cf \eqref{eq:path}. Instead, one could also consider limits for which different saxions $s^{k_i}$ scale at different rates 
\begin{equation}\label{eq:growthsector}
    s^{k_i} = e^{k_i} \lambda^{\alpha_{k_i}}\,. 
\end{equation}
Without loss of generality, we consider the case for which the vector $(\alpha_{k_1}, \dots, \alpha_{k_n})$ is ordered such that $\alpha_{k_i} > \alpha_{k_{i+1}}$.\footnote{We can also allow for $\alpha_{k_i}=\alpha_{k_{i+1}}$ but this just effectively reduces the number of enhancement steps that we consider in the following.} Consider a codimension-one singularity $\Delta_{k_i}\subset \cM_{\rm c.s.}(V)$ defined in suitable local coordinates as $u^{k_i}=0$, corresponding to $s^{k_i}\to \infty$ while keeping all other saxions constant. Then the path~\eqref{eq:growthsector} can be viewed as approaching the singularity at $\Delta_{k_1,\dots, k_n}$ by moving from a generic point in $\cM_{\rm c.s.}$ to a generic point on the singular divisor $\Delta_{k_1}$ and continuing further to a generic point on the intersection $\Delta_{k_1,k_2}$ and so on. With this path in the moduli space we can associate a chain of enhancements of singularity types, 
\begin{equation}\label{enhancements}
    {\rm I}_0\to \text{type}_{a_1}(\Delta_{k_1})\to \text{type}_{a_2}(\Delta_{k_1 k_2})\to\cdots \to \text{type}_{a_n}(\Delta_{k_1\cdots k_n})\,,
\end{equation}
where type$_{a_r}(\Delta_{k_1\cdots k_r})$ is the singularity type along the codimension-$r$ singularity $\Delta_{k_1,\dots, k_r}$, with $a_r = i^{2,2}$ denoting the secondary singularity type.  The possible enhancements have been discussed in~\cite{Kerr2017,Grimm:2018cpv}. 

We can distinguish two classes of enhancements: $i)$ the primary enhancement type changes and $ii)$ the primary enhancement type does not change. 

\subsection{Enhancement of Primary Singularity Type}\label{subsec:primary}

As for class $i)$, one can distinguish four cases, for which we  give a geometric and physical interpretation. For simplicity, we focus on the first enhancement step in \eqref{enhancements} by moving from $\Delta_{k_1} \to \Delta_{k_1,k_2}$.

 Even before coming to the concrete enhancements, we can anticipate an important general rule:
 The species scale in such enhancements is always determined, to leading order, by the scaling of the saxions responsible for the first infinite distance limit.
 To see this, we notice that in the vector multiplet moduli space of 4d $\cN=2$ theories the species scale can be estimated by the coefficient, $F_1$, of the Gauss-Bonnet term~\cite{vandeHeisteeg:2022btw,vandeHeisteeg:2023ubh} as 
\begin{equation}
    \frac{\Lambda_{\rm sp}}{M_{\rm Pl}} \sim \frac{1}{\sqrt{F_1}}\,. 
\end{equation}
 In asymptotic limits in which a continuous shift symmetry is realized, the coefficient of the Gauss-Bonnet term is at best linear in the moduli (up to exponential corrections)~\cite{Martucci:2024trp} 
\begin{equation}
    F_1 \sim \sum_{\alpha} f_{i} t^{i} \,. 
\end{equation}
For the limits discussed in this section this results in 
\begin{equation} \label{specie-pred}
    \frac{\Lambda_{\rm sp}}{M_{\rm Pl}} \sim \frac{1}{\sqrt{f_{k_1}\lambda}}\left(1 - \frac12 \frac{f_{{k_2}}}{f_{{k_1}}\lambda^{1-\alpha_{k_2}}} + \dots \right) \sim \lambda^{-1/2} + \mathcal{O}(\lambda^{-3/2+\alpha_{k_2}}) \,. 
\end{equation}
Thus, taking the sub-leading limit does not change the overall scaling of the species scale. This reasoning holds for any kind of enhancement. 
 Furthermore, note that the question how to determine the species scale in the presence of several towers was previously considered systematically in \cite{Castellano:2021mmx} from the point of view of counting the tower species.

Let us now see how this general prediction is indeed realized explicitly. This question was also addressed from the dual Type IIA perspective for concrete two-parameter examples in \cite{Castellano:2023jjt}, and the transition between asymptotic duality frames depending on the saxionic scalings is studied in various setups in \cite{Etheredge:2024tok,Grieco:2025bjy}.

\paragraph{Type II $\to$ III:} In this case the singular variety along $\Delta_{k_1}$ is a normal crossing variety $V_0$ for which the components intersect pairwise over a K3 surface $Z$\footnote{Alternatively, when $b=2$, $Z$ can also be a 4-torus \cite{Friedrich:2025gvs}, corresponing to an emergent string limit with a critical Type II string. The discussion for K3 surfaces easily carries over, mutatis mutandis, to such limits.} with transcendental lattice $\Gamma_{\rm trans}$ of rank $(2,b)$. As explained in detail in \cite{Friedrich:2025gvs}, locally a 1-cycle shrinks over $Z$ and combines with the 2-cycles on $Z$ that are not inherited from the components $V_i$ into a set of vanishing 3-cycles in ${\rm Gr}_2(\Delta_{k_1})$.

Enhancements to type III are possible only when $b\geq 2$. Such enhancements correspond to a further degeneration in which the K3 $Z$ itself undergoes a type II degeneration in the Kulikov classification \cite{Kulikov1977, Kulikov1981}, i.e., 
    \begin{equation}
        Z \to Z_0 = Z_1 \cup_{\cE} Z_2 \cup_{\cE} \dots \cup_{\cE} Z_n\,. 
    \end{equation}
As indicated in this expression, the components intersect pairwise over elliptic curves $\cE$. In this limit, there are two special Lagrangian 2-tori inside $Z$ that shrink to zero size, corresponding to the sublattice $\Gamma_{2,2}\subset \Gamma_{\rm trans}$ of signature $(2,2)$ of the transcendental lattice of $Z$. 
These 2-cycles are vanishing 2-tori and arise by fibering a vanishing 1-cycle over the two 1-cycles in ${\cal E}$. 
Now, the two 3-tori, that are the hallmark of type III degenerations, are obtained by fibering the vanishing 1-cycle normal to $Z$ that occurs in the initial type II limit of $V_0$ over these two 2-tori.   
This geometric picture gives a simple explanation for why enhancements from type II to type III are possible only starting from II$_{b\geq 2}$ enhancements. This fact was previously known also from a Hodge theory point of view \cite{Kerr2017,Grimm:2018cpv}.

We claim that the type II $\to$ III enhancement is an emergent string limit in six dimensions. To see this, recall from~\cite{Friedrich:2025gvs} that a type II$_b$ limit with $b\geq 2$ corresponds to the weak coupling limit of an emergent heterotic string compactified on ${\rm K3}\times T^2$. We parametrize the limit approaching the divisor $\Delta_{k_1}$ as 
    \begin{equation}\label{scalingtypeII}
        s^{k_1} = e^{k_1} \lambda\,, 
    \end{equation}
and recall from (\ref{EFT-tension-limit}) that the heterotic string scale (and thus the species scale) is set by  
    \begin{equation}\label{M_het}
        \frac{M_{\rm het}^2}{M_{\rm Pl}^2} = \frac{\Lambda_{\rm sp}^2}{M_{\rm Pl}^2} \sim \lambda^{-1}\,. 
    \end{equation}
Approach now the intersection $\Delta_{k_1,k_2}$ by also sending 
    \begin{equation}\label{scalingsk2}
        s^{k_2} = e^{k_2} \lambda^{\alpha_{k_2}}\,,\qquad \a_{k_2}<1\,.
    \end{equation}
From the heterotic perspective, this corresponds to scaling the radii of the two circles in $T^2 \simeq S^1 \times S^1$ to infinity in heterotic string units, 
    \begin{equation}
       r_{S^1} M_{\rm het} \sim \lambda^{\alpha_{k_2}/2}\to \infty \,.
    \end{equation}
This can be seen by noting that the KK states which become light through the enhancement to type III come from D3-branes with charge vector 
    $q \in {\rm Gr}_2(\Delta_{k_1}) \cap {\rm Gr}_1(\Delta_{k_1,k_2})$.
With the help of the growth theorem \eqref{eq:growththeorem}, their associated KK mass is seen to scale as 
    \begin{equation}
        \frac{M_{\text{KK}}^2}{M^2_{\rm Pl}}\sim|| q ||^2 \sim \left(\frac{s^{k_1}}{s^{k_2}}\right)^{-1} \left(s^{k_2}\right)^{-2} = \frac{1}{\lambda \, \lambda^{\alpha_{k_2}}} \,,
    \end{equation}
so that
    \begin{equation}
        \frac{M_{\rm KK}}{M_{\rm het}}\sim \lambda^{-\alpha_{k_2}/2}\,.
    \end{equation}    
The fact that the mass scale associated with the two KK towers from the type III limit lies parametrically below the heterotic string scale signals a decompactification to six dimensions even in the heterotic duality frame. 
Importantly, however, the net effect is not just 
a decompactification limit to six dimensions with all information about the heterotic string lost, but rather an emergent string limit in six dimensions. To determine the difference between both options, we must ask whether the heterotic string scale lies above or below the six-dimensional Planck scale~\cite{Castellano:2021mmx}. The latter is the species scale for the decompactification from 4d to 6d, which is related to the KK scale as in (\ref{5d6dspeciesscales}). In the limit, we find for this quantity
    \begin{equation}
        \left(\frac{M_{\rm KK}}{M_{\rm Pl}}\right)^{\frac12} \sim \frac{1}{(\lambda \lambda^{\alpha_{k_2}})^{1/4}} \gg \frac{1}{\lambda^\frac12} \sim \frac{M_{\rm het}}{M_{\rm Pl}} \quad \text{for} \quad \l\to \infty \,,\alpha_{k_2}<1\,. 
    \end{equation}
This shows that the heterotic string scale is below the six-dimensional Planck scale unless $\alpha_{k_2} =1$ and hence it is $M_{\rm het}$  which sets the true species scale in six dimensions, in agreement with (\ref{specie-pred}).

Thus the limit is an emergent string limit for which the compactification manifold of the emerging string decompactifies to six dimensions. The asymptotic theory is thus a weakly coupled heterotic string on K3. By contrast, for $\alpha_{k_2}=1$, the species scale is indeed given by the scale computed for a double KK tower consistent with our expectation that for an EFT limit ($\alpha_{k_2}=1$) the type III limit is a decompactification limit to 6d. In particular, in this case the heterotic string scale sits at the 6d Planck scale and therefore does not become tensionless from the point of view of the 6d EFT. 
 \paragraph{Type II $\to$ IV:} If the K3-surface $Z$ arising at a type II$_{b}$ singularity, $b\geq 1$, undergoes a type III Kulikov degeneration, the type II singularity enhances to a type IV singularity. In this case, the K3 surface $Z$ with transcendental lattice of signature $(2,b)$ with $b\geq 1$ degenerates
 \begin{equation}
     Z\to Z_0 = \bigcup_{i=1}^n Z_i\,,
 \end{equation}
 where some triple intersections are non-empty, $Z_{ijk}= Z_i\cap Z_j\cap Z_j\neq \emptyset$. Similar to the type IV singularities of Calabi--Yau threefolds there is now a single $T^2$ shrinking to zero size corresponding to a sublattice $\Gamma_{1,1}\subset \Gamma_{\rm trans}$ of signature $(1,1)$ inside the transcendental lattice of $Z$. Notice that for a Kulikov type III degeneration we only require a sublattice of signature $(1,1)$ as opposed to the sublattice with signature $(2,2)$ required for type II Kulikov degeneration. This is the geometric reason why for an enhancement type II$_b \to$ IV we only require $b\geq 1$ as opposed to $b\geq 2$ for a type II$_b$ $\to$ III enhancement. Together with the $S^1$ arising from the normal bundle of $Z$ inside $V_0$ the single shrinking $T^2$ inside $Z_0$ forms a single $T^3$ that we expect for a type IV singularity. \newline 

Physically, the type II $\to$ IV enhancement corresponds to an emergent string limit in five dimensions. As before, the type II singularity along the divisor $\Delta_{k_1}$ corresponds to an emergent string limit as in \eqref{scalingtypeII} and \eqref{M_het}. We can now approach the type IV singularity along $\Delta_{k_1,k_2}$ as in~\eqref{scalingsk2}. For the heterotic string on ${\rm K3}\times T^2$ this corresponds to only scaling the radius of one of the circles in $T^2\simeq S^1 \times S^1$ to infinity at a rate 
   \begin{equation}
       r_{S^1} M_{\rm het} \sim \lambda^{\alpha_{k_2}}\to \infty \,.
    \end{equation}
Again this can be seen using the growth theorem and that the KK tower has charge $q\in \text{Gr}_2(\Delta_{k_1})\cap \text{Gr}_0(\Delta_{k_1,k_2})$. As in the previous case we can compute from (\ref{5d6dspeciesscales}) the would-be species scale for this single KK tower as
    \begin{equation}
                \left(\frac{M_{\rm KK}}{M_{\rm Pl}}\right)^{\frac13} \sim \frac{1}{\l^\frac16 \lambda^{\frac{\alpha_{k_2}}{3}}} \gg \frac{1}{\lambda^\frac12} \sim \frac{M_{\rm het}}{M_{\rm Pl}} \quad \text{for} \quad \l\to \infty \,,\alpha_{k_2}<1\,.
    \end{equation}
For $\alpha_{k_2}<1$ this is above the actual species scale set by the heterotic string scale, again confirming (\ref{specie-pred}). Thus the limit is indeed an emergent string limit for which the compactification manifold of the emerging string decompactifies to five dimensions. Asymptotically we thus obtain the weakly coupled heterotic string on K3$\times S^1$. 
\paragraph{Type III $\to$ IV:} The geometry of this enhancement was already discussed in Section~\ref{subsec:III}. In this case the torus arising as the triple curve at a type III degeneration undergoes a further degeneration thereby inducing a hierarchy between the volume of the two $T^3$s shrinking at the type III degeneration. \newline

Consider a type III singularity along a divisor $\Delta_{k_1}$ which we approach as in \eqref{scalingtypeII}. Along $\Delta_{k_1}$ we have two KK towers with charge $q\in {\rm Gr}_1(\Delta_{k_1})$ becoming light at the rate 
\begin{equation}
    \frac{M_{\rm KK, 4d \to 6d}}{M_{\rm Pl}} \sim \lambda^{-1} \,,
\end{equation}
such that the 6d Planck scale (the species scale in this case) is given by 
\begin{equation}\label{speciestypeIII}
    \frac{\Lambda_{\rm sp, 4d \to 6d}}{M_{\rm Pl}} = \left(\frac{M_{\rm KK, 4d \to 6d}}{M_{\rm Pl}}\right)^{1/2}\sim \lambda^{-1/2} \,.
\end{equation}
Consider now the intersection of the divisors $\Delta_{k_1,k_2}$ corresponding to a type IV singularity which we approach by imposing the additional scaling~\eqref{scalingsk2}. In this case the radii of the two $S^1$s decompactifying in the original type III limit scale as 
\begin{equation}
    r_1 M_{\rm Pl}\sim (\lambda^2 \lambda^{\alpha_{k_2}})^{1/2} \,,\qquad \frac{r_1}{r_2} \sim \lambda^{\alpha_{k_2}}\gg 1\,. 
\end{equation}
Indeed, the two towers of states associated with the type III degeneration split up into a lighter and heavier one.
The leading tower comes from states with charges $q \in {\rm Gr_1}(\Delta_{k_1}) \cap {\rm Gr_0}(\Delta_{k_1,k_2})$ with associated mass scale, according to the growth theorem, given by
\begin{equation}
\frac{(M^{(1)}_{\rm KK})^2}{M^2_{\rm Pl}} \sim \lambda^{-2 -\alpha_{k_2}} \,.
\end{equation}
The heavier tower, on the other hand, has charges
$q \in {\rm Gr_1}(\Delta_{k_1}) \cap {\rm Gr_2}(\Delta_{k_1,k_2})$
and 
\begin{equation}\label{eq:MKK2} 
\frac{(M^{(2)}_{\rm KK})^2}{M^2_{\rm Pl}} \sim \lambda^{-2 +\alpha_{k_2}} \,.
\end{equation}
As we now argue, the resulting limit is still a decompactification to six dimensions. To that end, let us \emph{assume} that the limit was a decompactification to five dimensions. Then the species scale would be determined by the single tower with mass $M^{(1)}_{\rm KK}$ as 
\begin{equation}
    \left(\frac{M_{\rm KK}^{(1)}}{M_{\rm Pl}}\right)^\frac13 \sim \frac{1}{(\lambda^2 \lambda^{\alpha_{k_2}})^{1/6}}\,. 
\end{equation}
Comparing with \eqref{eq:MKK2}, we observe that for $\alpha_{k_2}<1$, the scale of the second KK tower, $M^{(2)}_{\rm KK}$ is below this would-be species scale such that there indeed exists a second KK tower signaling a further decompactification to 6d. 

Notice once more that, even for $\alpha_{k_2}<1$, the species scale of the decompactification to 6d is to leading order given by $\lambda^{-1/2}$ as in \eqref{speciestypeIII}, in agreement with the leading behaviour predicted by (\ref{specie-pred}).

\paragraph{Type II$\to$ III $\to$ IV:}
Finally, one can also consider a three-fold enhancement chain $\Delta_{k_1} \to \Delta_{k_1,k_2} \to \Delta_{k_1,k_2,k_3}$. The starting point for the enhancement is a type II$_{b \geq 2}$ singularity along 
$\Delta_{k_1}$, exhibiting an emergent string with associated string scale ${M_{\rm het}}$ as in (\ref{M_het}). 
The \emph{leading} KK tower is given by the states with $q^{(1)}\in {\rm Gr}_{2}(\Delta_{k_1})\cap {\rm Gr}_1(\Delta_{k_1,k_2}) \cap {\rm Gr}_0(\Delta_{k_1,k_2,k_3}) 
 $, corresponding to D3-branes multi-wrapping the single 3-torus that shrinks fastest in the final type IV degeneration.
These KK states have mass
\begin{equation}
    \frac{(M^{(1)}_{\rm KK})^2}{M^2_{\rm Pl}}\sim \|q^{(1)}\|^2\sim \frac{1}{\lambda^{1+\alpha_{k_2}+\alpha_{k_3}}} 
\end{equation}
and a would-be species scale 
\begin{equation}
\left(\frac{M^{(1)}_{\rm KK}}{M_{\rm Pl}}\right)^{\frac{1}{3}}\sim \frac{1}{\lambda^{\frac{1}{6}(1+\alpha_{k_2}+\alpha_{k_3})}}\gg \frac{1}{\lambda^{\frac{1}{2}}} \sim \frac{M_{\rm het}}{M_{\rm Pl}}\,,\quad \text{for} \quad \l\to \infty \,,\alpha_{k_2}+\alpha_{k_3}<2\,.
\end{equation}
Again, there is also a subleading KK tower from states with 
 $q^{(2)}\in {\rm Gr}_{2}(\Delta_{k_1})\cap {\rm Gr}_1(\Delta_{k_1,k_2}) \cap {\rm Gr}_2(\Delta_{k_1,k_2,k_3}) 
 $
 with  mass scale
 \begin{equation}
\frac{M^{(2)}_{\rm KK}}{M_{\rm Pl}} \sim \frac{1}{\lambda^{\frac{1}{2}(1+\alpha_{k_2}-\alpha_{k_3} )  }} \,.
\end{equation}
For $1 > \alpha_{k_2} > \alpha_{k_3}$, 
 this results in the hierarchy (for $\lambda\to \infty$)
\begin{equation}
\label{eq:species-hierarchy}
\underbrace{ \lambda^{-\frac{1}{2}(1+\alpha_{k_2}+\alpha_{k_3} ) }}_{\sim M^{(1)}_{\rm KK}}  \ll \underbrace{\lambda^{-\frac{1}{2}(1+\alpha_{k_2}-\alpha_{k_3} ) }}_{\sim M^{(2)}_{\rm KK}}  \ll \underbrace{\lambda^{-1/2}}_{\sim M_{\rm het}}\ll \underbrace{\lambda^{-1/4(1+\alpha_{k_2})}}_{\sim (M_{\rm KK}^{(0)}/M_{\rm Pl})^{1/2}}  \ll \underbrace{ \lambda^{-\frac{1}{6}(1+\alpha_{k_2}+\alpha_{k_3} )}}_{\sim \left(M^{(1)}_{\rm KK}/M_{\rm Pl}\right)^{1/3}} \,,
\end{equation}
 where we have included the would-be species scale associated with the decompactification to six dimensions arising at the type III singularity for $\alpha_{k_3}=0$ with KK scale $M_{\rm KK}^{(0)}$. The above hierarchy shows that for $1>\alpha_{k_2}>\alpha_{k_3}$ the limit is an emergent string limit in which the spacetime decompactifies to six dimensions. Notice that for $\alpha_{k_2}=1$, the scale $M_{\rm het}$ coincides with the would-be 6d species scale signaling that we obtain a 6d limit without perturbative critical string. Similarly, for $\alpha_{k_2}=\alpha_{k_3}$ the scale $M_{\rm KK}^{(2)}$ coincides with $M_{\rm het}$ indicating that in this case we have an emergent string limit in which the spacetime decompactifies only to five dimensions. 
Both observations are consistent with the general rule that if some of the $\alpha_{k_i}$ coincide, one can combine the associated enhancements into one EFT string limit and consider it as a single block in the treatment of the remaining enhancement steps.

\begin{figure}
    \centering
    \includegraphics[width=\linewidth]{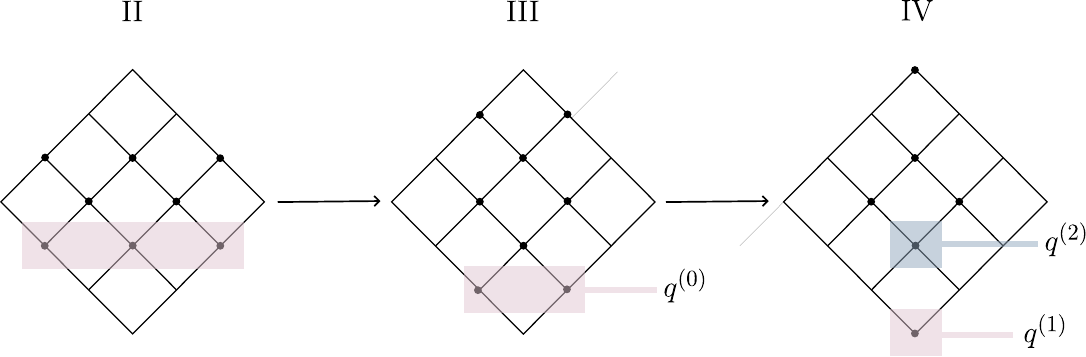}
    \caption{The limiting mixed Hodge structures for a $\rm{II}\to\rm{III}\to\rm{IV}$ enhancement chain. Along each step in the chain, the states indicated in pink correspond to the lightest towers and subsequently determine the would-be species scales as summarised in equation \eqref{eq:species-hierarchy}. In addition, the states in blue comprise the subleading KK tower.}
    \label{fig:enhancement}
\end{figure}

\paragraph{General enhancements:} The above four cases in fact cover all possibilities. To see this explicitly, consider an arbitrary enhancement chain
\begin{equation}
    {\rm I}_0\to \text{type}_{a_1}(\Delta_{k_1}) \to \text{type}_{a_2}(\Delta_{k_1,k_2}) \to \dots \to \text{type}_{a_n}(\Delta_{k_1,k_2,\dots,k_n})\,.
\end{equation}
For a limit approaching the codimension $h^{2,1}-n$ locus $\Delta_{k_1\cdots k_n}$, we introduce the would-be species scale $\tilde{\Lambda}_{\rm sp}$ associated with the lightest state corresponding to
\begin{equation}
    q \in \mathrm{Gr}_{3-d_1}(\Delta_{k_1})\cap\cdots\cap\mathrm{Gr}_{3-d_{k_1\cdots k_n}}(\Delta_{k_1\cdots k_n})\,.
\end{equation}
Then one finds
\begin{equation}
    \frac{\tilde{\Lambda}_{\rm sp}}{M_{\rm Pl}} = \left(\frac{M_q}{M_{\rm Pl}}\right)^{\frac{D-4}{D-2}}\sim \lambda^{-\frac{1}{2}\frac{D-4}{D-2}\sum_{i=1}^n \alpha^i \left(d_{k_1\cdots k_{i+1}}-d_{k_1\cdots k_i}\right)}\,,
\end{equation}
with $D=6,5, \infty$ for decompactifications to $6$ and $5$ dimensions and emergent string limits, respectively.
Importantly, due to the presence of the difference $d_{k_1\cdots k_{i+1}}-d_{k_1\cdots k_i}$ the would-be species scale remains the same across limits of the same primary type. Of course, the same holds true for the scaling of the mass of the leading and sub-leading towers. Hence, for the purpose of understanding the emerging gravitational theory along a given enhancement chain, it suffices to consider only enhancements of the primary singularity type, for which we have discussed all possibilities. Let us also remark that one may alternatively write
\begin{equation}
    \frac{\tilde{\Lambda}_{\rm sp}}{M_{\rm Pl}} \sim \lambda^{-\frac{1}{2}\frac{D-4}{D-2}\sum_{i=1}^n (\alpha^i-\alpha^{i+1})d_{k_1\cdots k_{i+1}}}\,.
\end{equation}
This makes apparent the fact that the would-be species scale does not change when subsequent singularities are approached at the same rate. The EFT string limits are the extreme case in which all singularities are approached at the same rate, such that the would-be species scale is the same across the whole enhancement chain. Instead, for non-EFT string limits there is the hierarchy summarised in \eqref{eq:species-hierarchy}. 

\begin{figure}
    \centering
    \includegraphics[width=\linewidth]{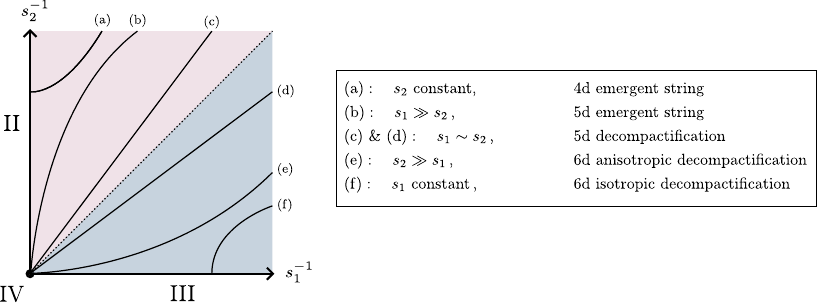}
    \caption{Illustrated are a type $\rm{II}$ divisor $(s_1\to\infty)$ and type $\rm{III}$ divisor ($s_2\to \infty$) intersecting in a type $\rm{IV}$ singularity. The two growth sectors corresponding to the $\rm{II}\to\rm{IV}$ ($s_1>s_2$) and $\rm{III}\to\rm{IV}$ $(s_2> s_1)$ enhancements are depicted in pink and blue, respectively. Within each growth sector we have indicated the qualitatively different trajectories $\mathrm{(a)}-\mathrm{(f)}$, whose physical interpretation is listed on the right.}
    \label{fig:example-enhancement}
\end{figure}

\paragraph{Summary \& generalisation to other trajectories.} The discussion of the different enhancements of the primary singularity type reveals that for an enhancement chain of the form~\eqref{enhancements}, the emerging gravitational theory at the end-point of the limit depends on the chosen trajectory \eqref{eq:growthsector}, but in a controlled way. There are two cases to consider:
\begin{itemize}
    \item \textbf{Case (A): non-EFT string limit}\\
    If there is an asymptotic hierarchy between two saxions, $s^{k_1}\gg s^{k_2}$, then the primary singularity type of the \emph{first} singularity in the chain, i.e. at $\Delta_{k_1}$, determines the physical interpretation of the limit. In particular, the physics associated with this singularity determines the species scale (see (\ref{specie-pred})) and hence the quantum gravitational duality frame emerging. The subsequent enhancements of the primary singularity type then merely tell us whether within this duality frame an additional limit is taken. For example, along such a trajectory the $\rm{II}\to\rm{III}$ enhancement corresponds to an emergent string limit in 6d. Furthermore, as illustrated in the $\rm{II}\to\rm{III}\to\rm{IV}$ case, if there are multiple subsequent enhancements then one should again distinguish whether $s^{k_2}\gg s^{k_3}$ or $s^{k_2}\sim s^{k_3}$ as discussed in Case (B). 
    \item \textbf{Case (B): EFT string limit}\\
    If the two saxions scale at the same rate, $s^{k_1}\sim s^{k_2}$, then it is the primary singularity type of the \emph{second} singularity in the chain, i.e. at $\Delta_{k_1k_2}$, that determines the physical interpretation of the limit. For example, along such a trajectory the $\rm{II}\to\rm{III}$ enhancement corresponds to a pure decompactification limit to 6d. 
\end{itemize}
See also Figure \ref{fig:example-enhancement} for an overview of the various interpretations for the example of a $\langle\rm{II}|\rm{IV}|\rm{III}\rangle$ limit. 

For simplicity of exposition, we have centred the above discussion around polynomial trajectories in the moduli space of the form \eqref{eq:growthsector}. One could, in principle, also consider more general trajectories that, for example, feature also exponential hierarchies (taking $s^1\sim\lambda$ and $s^2\sim \log\lambda$). Based on the above observations, we expect that the physical interpretation along general trajectories follows the same pattern. Namely, the only relevant feature is whether two saxions scale at the same rate or not. 

\subsection{Enhancement of Secondary Singularity Type}
Whereas the enhancement of the primary singularity type encodes the details of the \emph{gravitational} theory emerging at the singular locus in the moduli space, the secondary singularity type encodes information about the field theory sectors that couple/decouple from gravity.\footnote{In  \cite{Marchesano:2024tod} the relation between enhancement chains, decoupled field theory sectors and the moduli space curvature in Type IIA Calabi--Yau threefold compactifications is discussed briefly, and more generally the role of decoupled field theory sectors in infinite distance limits is discussed in \cite{Marchesano:2023thx,FierroCota:2023bsp,Castellano:2024gwi,Blanco:2025qom}. } Consider for simplicity an enhancement of the form  
\begin{equation}\label{enhancementsecondary}
    \text{type}_{b_1}(\Delta_{k_1}) \to \text{type}_{b_2}(\Delta_{k_1k_2})\,,
\end{equation}
realized by moving along a divisor $\Delta_{k_1}$ towards the intersection with a second divisor $\Delta_{k_2}$. In addition, suppose that the primary enhancement type remains the same and only the secondary enhancement type changes, $b_1\neq b_2$. In terms of the dimensions of the space $I^{p,q}$ this means that all $i^{3,q}$ remain the same and $i^{2,2}$ changes by 
\begin{equation}
\delta i^{2,2} = b_2-b_1 > 0 \,.
\end{equation}
Since $i^{3,q}$ does not change, the nature of the emerging gravitational theory remains the same. 

Rather than affecting the gravitational sector, an enhancement of the secondary enhancement type signals that a $\mathrm{U}(1)^{\delta i^{2,2}}$ gauge theory, which is originally strongly coupled and decoupled from gravity, becomes coupled to gravity. To see this, we introduce the scale 
\begin{equation}
    \Lambda_{{\rm WGC},q}^2 = g_{\mathrm{U}(1)_q}^2 M_{\rm Pl}^2\,. 
\end{equation}
Here, $g_{\mathrm{U}(1)_q}^2$ is defined in the following way: Consider a basis of $\mathrm{U}(1)$ gauge factors of the theory of gravity. A given state with charge $q$ in this basis then specifies a linear combination of the $\mathrm{U}(1)$ factors which we denote by $\mathrm{U}(1)_q$. The gauge coupling of this $\mathrm{U}(1)_q$ is then given by the physical charge of the state $q$, i.e. 
\begin{equation}
    g_{\mathrm{U}(1)_q}^2 = \|q\|^2\,. 
\end{equation}
As discussed in \cite{FierroCota:2023bsp,Marchesano:2024tod} a $\mathrm{U}(1)_q$ gauge theory can be considered to be decoupled from gravity if 
\begin{equation}
    \frac{\Lambda_{{\rm WGC},q}}{\Lambda_{\rm sp}} \gg 1\,. 
\end{equation}
To determine whether a $\mathrm{U}(1)_q$ gauge theory is decoupled from gravity, we hence have to consider the physical charge of a state with charge $q$ and compare it to the species scale.

We can follow this recipe, taking for concreteness the scalings as in~\eqref{scalingtypeII} and \eqref{scalingsk2}.
\begin{figure}
    \centering
    \includegraphics[width=0.6\linewidth]{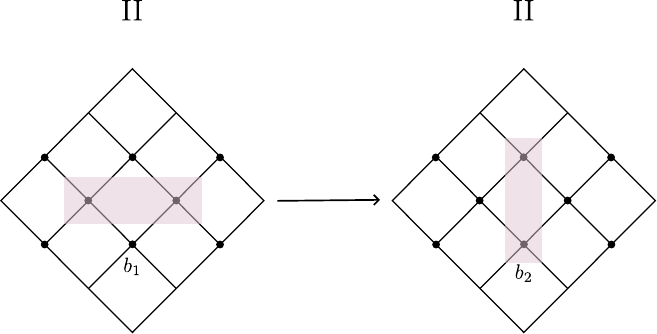}
    \caption{An enhancement $\mathrm{II}_{b_1}\to\mathrm{II}_{b_2}$, with $b_2>b_1$, that only involves a change in the secondary singularity type. The change is induced by $\delta i^{2,2}=b_2-b_1$ states in $\mathrm{Gr}_3(\Delta_{k_1})$ on the left-hand side moving to $\mathrm{Gr}_2(\Delta_{k_1 k_2})$ and $\mathrm{Gr}_4(\Delta_{k_1 k_2})$ on the right-hand side.}
    \label{fig:enhancement-secondary}
\end{figure}
 The primary type in this case is type ${\rm II}$, and along the enhancement chain~\eqref{enhancementsecondary}  a total of $\delta i^{2,2}$ states from $\text{Gr}_{3} (\Delta_{k_1})$ change their Hodge type and end up as new states in 
$\text{Gr}_{2} (\Delta_{k_1, k_2})$, which thereby increases by $\delta i^{2,2}$.\footnote{The same holds true for $\rm{III}\to\rm{III}$ and $\rm{IV}\to\rm{IV}$ enhancements.}
This means that 
\begin{equation}
{\rm dim}(\text{Gr}_3(\Delta_{k_1}) \cap \text{Gr}_2(\Delta_{k_1,k_2})) = \delta i^{2,2} \,.
\end{equation}
Using the growth theorem, we find for these states 
\begin{equation}
    \|q\|^2 \sim \left(\frac{s^{k_1}}{s^{k_2}}\right)^{0} \left(s^{k_2}\right)^{-1} \sim \lambda^{-\alpha_{k_2}}\,.  
\end{equation}
Thus the WGC scale of the associated $\mathrm{U}(1)_q$ is
\begin{equation}
    \frac{\Lambda_{{\rm WGC},q}}{M_{\rm Pl}} \sim \lambda^{-\frac{\alpha_{k_2}}{2}}\,. 
\end{equation}
From Section~\ref{subsec:primary}, we conclude that the scaling of the species scale is insensitive to the singularity type of the first singularity in the enhancement chain such that for $q\in \text{Gr}_3(\Delta_{k_1}) \cap \text{Gr}_2(\Delta_{k_1,k_2})$
\begin{equation}
    \frac{\Lambda_{{\rm WGC},q}}{\Lambda_{\rm sp}} = \lambda^{\frac{1-\alpha_{k_2}}{2}} \,. 
\end{equation}
Since there are $\delta i^{2,2}$ linearly independent such charges $q$, there is a $\mathrm{U}(1)^{\delta i^{2,2}}$ subgroup of the full $\mathrm{U}(1)^{h^{2,1}}$ gauge group for which the coupling to/ decoupling from gravity is parametrized by $\alpha_{k_2}$. Accordingly, for  $\alpha_{k_2} <1$  the gauge theory associated with the $\mathrm{U}(1)^{\delta i^{2,2}}$ subgroup is decoupled from gravity and gets coupled to gravity for $\alpha_{k_2}\to 1$ corresponding to the EFT limit approaching $\Delta_{k_1,k_2}$ as in this case $\Lambda_{\rm WGC,q} \sim \Lambda_{\rm sp}$.

\paragraph{Including enhancements of primary singularity type.} To discuss how the above observations generalise, let us consider an $n$-parameter limit with $q\in \mathrm{Gr}_{3+\ell_1}(\Delta_{k_1})\cap \cdots \cap  \mathrm{Gr}_{3+\ell_n}(\Delta_{k_1 \cdots k_n})$. Then we have
\begin{equation}
    \frac{\Lambda_{\mathrm{WGC},q}}{\Lambda_{\rm sp}} \sim \lambda^{\frac{1}{2}\left(1+\sum_{i=1}^n \ell_i\,\Delta\alpha_{k_i}\right)}\,,\qquad \Delta\alpha_{k_i} = \alpha_{k_i}-\alpha_{k_{i+1}}\,.
\end{equation}
In particular, the ``gravitational'' $\mathrm{U}(1)$ gauge groups, i.e.~those which remain coupled to gravity, are those for which $\Lambda_{\mathrm{WGC},q}\lesssim \Lambda_{\rm sp}$, which translates into the condition
\begin{equation}
\label{eq:grav-condition}
    1+\sum_{i=1}^n \ell_i\,\Delta\alpha_{k_i}\leq 0\,.
\end{equation}
For EFT string limits (where $\alpha_{k_i}=1$ for all $i=1,\ldots, n$) this reduces to the condition $\ell_n\leq -1$. In other words, \emph{any} state that ends up in $\mathrm{Gr}_0(\Delta_{k_1\cdots k_n})\oplus \mathrm{Gr}_1(\Delta_{k_1\cdots k_n})\oplus \mathrm{Gr}_2(\Delta_{k_1\cdots k_n})$ gives rise to a gravitational $\mathrm{U}(1)$ gauge group. Hence, in an $n$-parameter EFT string limit: 
\begin{equation}\begin{aligned}\label{eq:gravicoupledU1s}
    \text{\# gravitational $\mathrm{U}(1)$ gauge fields} &= \mathrm{dim}\left[\mathrm{Gr}_0(\Delta_{k_1\cdots k_n})\oplus \mathrm{Gr}_1(\Delta_{k_1\cdots k_n})\oplus \mathrm{Gr}_2(\Delta_{k_1\cdots k_n})\right]\\
    &=\begin{cases}
        2+b\,,& \mathrm{II}_b\,,\\
        2+c\,,&\mathrm{III}_c\,,\\
        1+d\,,&\mathrm{IV}_d\,,
    \end{cases}
\end{aligned}\end{equation}
where the singularity type refers to that of the final singularity $\Delta_{k_1\cdots k_n}$ that is reached in the limit. 
The result \eqref{eq:gravicoupledU1s} agrees with the analysis of rigid field theory sectors in \cite{Castellano:2024gwi} (see also \cite{Marchesano:2024tod}):
 The number of gravitationally coupled U(1) gauge fields in the limit is given by the number of all U(1)s minus the number of rigid field theory U(1)s except for Type III limits, where we must subtract one more U(1) which becomes the self-dual 2-tensor (rather than a 1-form gauge field) in the gravity multiplet of the asymptotic six-dimensional supergravity. Concretely, the number of rigid field theory U(1)s is given by~\cite{Castellano:2024gwi} 
 \begin{equation}
     r_{\rm rigid} = h^{2,1} - r_{\rm K} -\lambda \,,
 \end{equation}
 where $(r_{\rm K},\lambda)=(b,1)$, $(c+2,0)$, and $(d,0)$ for type II$_b$, III$_c$ and IV$_d$ limits, respectively, in agreement with~\eqref{eq:gravicoupledU1s}.

Recalling the maximum allowed values for the subindices that determine the secondary singularity type (see for example \cite{Grimm:2018cpv}), we find that along EFT string limits the maximal number of gravitational $\rm{U}(1)$s is given by
\begin{equation}
    \max\,\text{\# gravitational $\mathrm{U}(1)$ gauge fields} = \begin{cases}
        h^{2,1}+1\,,& \mathrm{II}_{h^{2,1}-1}\,,\\
        h^{2,1}\,,& \mathrm{III}_{h^{2,1}-2}\,,\\
        h^{2,1}+1\,,&\mathrm{IV}_{h^{2,1}}\,.
    \end{cases}
\end{equation}
In particular, we see that there always remains at least one decoupled $\mathrm{U}(1)$ gauge field in any limit that ends in a type III singularity. This gives another indication that it must always be possible to further enhance a type III singularity to a type IV singularity, as discussed in Section \ref{subsec:III}. Finally, let us remark that there can also be gravitational $\mathrm{U}(1)$s arising along trajectories that are not EFT string limits. For example, it is clear from \eqref{eq:grav-condition} that any state for which $\ell_i\leq -1$, for all $i=1,\ldots, n$, gives rise to a gravitational $\mathrm{U}(1)$, regardless of the precise trajectory. However, if a state has $\ell_i>0$ for some $i$, then the details of the trajectory become important in order to address the fate of the $\mathrm{U}(1)$. Intuitively, the more a trajectory deviates from being an EFT string limit, i.e.~the larger $\Delta\alpha_{k_i}$, the more difficult it is to satisfy the inequality \eqref{eq:grav-condition}. Thus, our expectation is that the maximal number of gravitational $\mathrm{U}(1)$s occurs along EFT string limits. However, a complete characterisation of the precise path-dependence is expected to be challenging, as already emphasised in \cite{Grimm:2018cpv,Grimm:2022sbl}.

\section{Discussion and Conclusions}
In this work, we have elucidated the physical interpretation of infinite-distance limits in the vector multiplet moduli space of Type IIB Calabi--Yau compactifications to four dimensions. Our analysis has supplemented previous works that focused on the algebraic properties of the low-energy effective theory with a geometric characterisation of infinite-distance limits. Based on an understanding of the geometry of the infinite distance limits we have indeed found complete agreement with the expectations of the Emergent String Conjecture~\cite{Lee:2019oct}. 

On the \emph{algebraic} side, we have focused on the asymptotic behaviour of the physical couplings along trajectories in the complex structure moduli space that are induced by the backreaction of EFT strings. In particular, by utilising the underlying limiting mixed Hodge structures, we have analysed the scaling of the mass of BPS particles, as well as the tension of the EFT strings, and isolated the characteristic scaling laws that are associated with emergent string / decompactification limits. Importantly, we have also made explicit the two missing pieces that are required in order to rigorously establish the emergent dual description in the various type II/III/IV limits in the moduli space:
\begin{enumerate}
    \item There indeed exist the appropriate \emph{tower(s)} of asymptotically light states. 
    \item The towers have the correct \emph{degeneracy} corresponding to their microscopic interpretation as either a tower of string oscillators or a KK tower. In particular, for KK towers the density of states ought to be constant. 
\end{enumerate}
To address these points, we have leveraged the \emph{geometric} interpretation of the aforementioned limits, working in the framework of 
a semi-stable degeneration of the Calabi--Yau threefold.
  In this setting, there is a clear correspondence between the algebraic limiting mixed Hodge structure and the geometric mixed Hodge structure associated with the central fibre of the degeneration. The latter is then described in terms of certain cohomology groups of the various components and their intersections arising in the degeneration, as encoded by the formula \eqref{eq:Gr-general}. Additionally, this intersection structure is conveniently captured by a simplicial complex known as the dual graph of the central fibre, and its dimension provides a simple geometric characterisation of the type II/III/IV classification of infinite-distance limits. 

Our first main result concerns the geometry and topology of the special Lagrangian 3-cycles whose Poincar\'e duals constitute the charges $q\in {\rm Gr}_{3-d}$, where $d=1,2,3$ for type II/III/IV limits. We have argued that, in the vicinity of the degeneration, the topology of the 3-cycle $\Gamma$ takes the form of a torus-fibration
\begin{equation}
    T^d \hookrightarrow \Gamma\to B_{3-d}\,,
\end{equation}
such that the $T^\ell$-fibre shrinks to zero size in the limit. 
 While for type II degenerations the nature of these shrinking cycles was already discussed in \cite{Friedrich:2025gvs},
we have noticed that in limits of type III and type IV, there arise two and, respectively, one type of vanishing cycle $\Gamma$ with the topology of a three-torus.
Our second main result concerns the counting of the BPS indices associated to multi-wrappings of $\Gamma$. 
 As a result of the toroidal topology of $\Gamma$,
multi-wrapped D3-branes indeed give rise to
two (in type III limits) or one (type IV) tower of Kaluza-Klein type.
  Using the resulting interpretation of the type IV limit as a decompactification limit to five dimensions, we have related the BPS index of multi-wrapped D3-branes to the BPS index of the massless, uncharged states of the asymptotic five-dimensional theory, allowing us to argue 
\begin{equation}
    \Omega_{\rm BPS}(n\Gamma) = -\chi(V)\,,\qquad \forall n\in\mathbb{Z}_{>0}\,.
\end{equation}
  The same result carries over to type III limits.
 Let us stress that this identification of the BPS index with the Euler characteristic of the threefold is based on the identification of the type IV limits as decompactifications to five dimensions. In particular, we did not derive the BPS index by studying the details of the moduli space of A-branes on $\Gamma$. In this way, our result can be viewed as a physics prediction for the geometry of Calabi--Yau threefolds highlighting once again that the interplay between physics and geometry yields new insights on both sides.  

Our third main result concerns the geometric and physical interpretation of singularity enhancements, particularly along paths in the moduli space that deviate from those that are induced by EFT strings, i.e.~non-linear paths. By carefully comparing the scale of the KK modes with the relevant species scale, we have found that for such paths the emerging gravitational theory at the end-point of the infinite-distance limit is dictated by the primary singularity type of the \emph{first} singularity in the enhancement chain, while subsequent enhancements beyond the first singularity indicate an additional limit within this duality frame. In contrast, for the path induced by EFT strings it is the primary singularity type of the \emph{last} singularity in the enhancement chain that determines the physical interpretation of the limit. Additionally, we have investigated the interpretation of the secondary singularity type and found that it governs the rank of the gauge group that is decoupled from gravity. Notably, this provides a physical interpretation of some of the enhancement rules that were obtained by purely algebraic means in \cite{Kerr2017,Grimm:2018cpv}. An interesting question, which we leave for future investigation, is whether one can obtain new enhancement rules from this physical perspective. In particular, it would be interesting to constrain the possible $n$-cubes of polarised limiting mixed Hodge structure in this way, complementing the representation-theoretic approach of \cite{Kerr2017}.

Finally, let us stress that for the physical limits we are interested in, the framework of semi-stable degenerations is sufficient. From a more mathematical point of view, however,
  the works of Gross--Siebert  study degenerations of Calabi--Yau manifolds beyond the semi-stable setting \cite{GrossSiebert2003}. Although these works mostly focus on the large complex structure limit in the context of mirror symmetry and the SYZ conjecture, it would be interesting to see to what extent similar methods can be applied to other kinds of limits in the complex structure moduli space and what effects, if any, they add from a physics point of view.

\subsubsection*{Acknowledgements}
We thank Lukas Kaufmann, Stefano Lanza,  Dieter L\"ust, Fernando Marchesano, Luca Melotti, and Guo-En Nian for discussions and Bj\"orn Hassfeld for collaboration on related work. 
 This work is supported in part by Deutsche Forschungsgemeinschaft under Germany’s Excellence Strategy EXC 2121 Quantum Universe 390833306, by Deutsche Forschungsgemeinschaft through a German-Israeli Project Cooperation (DIP) grant “Holography and the Swampland” and by Deutsche Forschungsgemeinschaft through the Collaborative Research Center 1624 “Higher Structures, Moduli Spaces and Integrability.”

\appendix 

\section{The Dual Graph and Gromov--Hausdorff Limits}
\label{app:dual-graph}

Throughout this work, we have focused on the algebro-geometric properties of a semi-stable degeneration $\mathcal{V}\to\mathbf{D}$ and observed in Section \ref{sec:II_III_IV} that the dual graph $\Pi(V_0)$ characterises a part of the mixed Hodge structure on $V_0$ via the relation \eqref{eq:Gr0-dualgraph}. In the following, we aim to elucidate the role of $\Pi(V_0)$ from a more differentio-geometric point of view. In other words, we take into account what happens to the \emph{metric} of the Calabi--Yau $V_z$ as we approach the degeneration at $z\to 0$. This is done by considering the so-called \emph{Gromov--Hausdorff limit}. For further references and applications in the context of mirror symmetry and the Gross--Siebert program, we refer the reader to \cite{GrossSiebert2003,Gross:2012} and references therein.

Given a sequence of compact metric spaces $(X_i, d_i)$, we say that this sequence converges in the Gromov--Hausdorff sense to
\begin{equation}
    (X_i, d_i) \stackrel{\text{GH}}{\longrightarrow} (X_\infty, d_\infty)\,,
\end{equation}
if for every $\epsilon>0$ there exist maps
\begin{equation}
    f_i:X_i\to X_\infty\,,\qquad g_i:X_\infty\to X_i\,,
\end{equation}
such that for all $i$
\begin{equation}
    \left|d_i(x,y) - d_\infty(f(x_i), f(y_i)) \right|<\epsilon\,,\qquad \forall x_i, y_i\in X_i\,,
\end{equation}
and similarly for $g_i$. 

We would now like to apply the above notion to a sequence $V_{z^i}$ of Calabi--Yau manifolds equipped with a Ricci-flat metric $g_{i}$ (which we normalize to have bounded diameter), and ask what the resulting Gromov--Hausdorff limit is. To gain some intuition, we consider two examples. 

\paragraph{Example (I): Elliptic Curve.} We consider a family of elliptic curves defined by
\begin{equation}
    E_\tau = \mathbb{C}/(\mathbb{Z}+\tau\mathbb{Z})\,,\qquad \tau = \frac{\log z}{2\pi i}
\end{equation}
and consider a sequence $z_i$ approaching 0 along the positive real axis. After an appropriate rescaling of the metric on $E_\tau$, the periods of the holomorphic 1-form are $\frac{1}{\log|z|}$ and $i$. In particular, the Gromov--Hausdorff limit of such a sequence of elliptic curves is the circle $\mathbb{R}/\mathbb{Z}$. 

\begin{figure*}[t!]
    \centering
    \begin{subfigure}[t]{0.5\textwidth}
        \centering
        \includegraphics{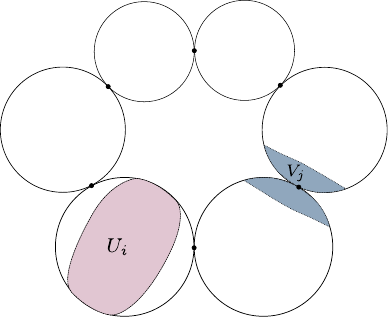}
        \caption{Semi-stable degeneration limit}
    \end{subfigure}%
    ~ 
    \begin{subfigure}[t]{0.5\textwidth}
        \centering
        \includegraphics{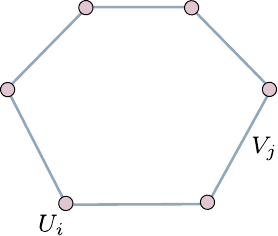}
        \caption{Gromov--Hausdorff limit}
    \end{subfigure}
    \caption{Two perspectives on the degeneration of an elliptic curve. In (a) the algebro-geometric perspective is depicted, in which the elliptic curve undergoes an I$_n$ singularity ($n=6$ in this example). In (b) the differentio-geometric perspective obtained via the Gromov--Hausdorff limit is depicted. In both figures the open sets $U_i$ and $V_j$ are depicted in pink and blue, respectively. Importantly, as explained in equations \eqref{eq:diam-U}--\eqref{eq:diam-V}, the diameter of the open set $V_j$ is constant in the Gromov--Hausdorff limit, while the diameter of the open set $U_i$ tends to zero. }
    \label{fig:GS}
\end{figure*}

\paragraph{Example (II): $I_n$ Singularity of an Elliptic Curve. }
To make contact with the more algebro-geometric point of view, let us again consider a sequence of elliptic curves as in the previous example such that $E_\tau$ acquires an $I_n$ singularity in the limit. This means that, after resolving the singularity, the central fibre $E_0$ is given by the union of $n$ $\mathbb{P}^1$'s
\begin{equation}
    E_0 = \mathbb{P}^1_1\cup\cdots\cup\mathbb{P}^1_n\,,
\end{equation}
which intersect pairwise in a chain. In this case the Gromov--Hausdorff limit will give the circle $\mathbb{R}/n\mathbb{Z}$. We would like to understand how the latter relates to the various components in the semi-stable degeneration. To this end, we follow the discussion in \cite{Gross:2012} and consider two kinds of open sets in the total space $\mathcal{V}$. 
\begin{enumerate}
    \item First, one may consider tubular sets $U_i$, each of which surrounding but \textit{not including} an intersection point $\mathbb{P}^1_i\cap\mathbb{P}^1_{i+1}$. This gives rise to an open subset $\tilde{U}_i$ of the total space which is biholomorphic to $U_i\times \mathbf{D}_\rho$, where the latter is a disk of some radius $\rho$. Then one can show that the diameter of $\tilde{U}_i$ as measured in the normalized metric (as in the previous example) scales as
    \begin{equation}
    \label{eq:diam-U}
        \mathrm{diam}(\tilde{U}_i)\sim \frac{1}{\log|z|}\stackrel{z\to 0}{\longrightarrow} 0\,.
    \end{equation}
    In particular, it shrinks to zero size as we approach the degeneration. 
    \item Second, one may consider a set $V_i$ which \textit{does} include the intersection point $\mathbb{P}^1_i\cap\mathbb{P}^1_{i+1}$. This gives rise to an open subset $\tilde{V}_i$ of the total space which is biholomorphic to a polydisk. Then one may show that the diameter of $\tilde{V}_i$ scales as
    \begin{equation}
    \label{eq:diam-V}
        \mathrm{diam}(\tilde{V}_i)\sim \mathcal{O}(1)\,.
    \end{equation}
    In particular, it approaches an interval whose length may be normalized to one. 
\end{enumerate}
The upshot of the above discussion is that, in the Gromov--Hausdorff sense, the individual $\mathbb{P}^1$-components (excluding the intersection points) shrink to a point, while the intersection points give rise to lines. In other words, when taking into account the degeneration of the metric, the ``asymptotic geometry'' $E_0$ is precisely described by the \emph{dual graph} of $E_0$. In the above example, the latter is simply an $S^1$. See also Figure \ref{fig:GS} for an illustration. 

Returning to the general case, the above examples have motivated the following conjecture by Gross--Wilson \cite{Gross:2001} and Kontsevich--Soibelman \cite{Kontsevich:2001}:
\begin{quote}
    \textbf{Conjecture:} Let $\mathcal{V}\to \mathbf{D}$ be a maximally unipotent degeneration of simply-connected Calabi--Yau manifolds with full $\mathrm{SU}(n)$ holonomy, $z_i\in\mathbf{D}$ with $z_i\to 0$, and let $g_i$ be a Ricci-flat metric on $V_{z^i}$ normalized to have fixed diameter. Then a convergent subsequence of $(V_{z^i}, g_i)$ converges to a metric space $V_\infty$, where $V_\infty$ is homeomorphic to the $n$-sphere $S^n$. 
\end{quote}
Let us stress here the restriction to points of maximally unipotent monodromy, i.e.~large complex structure points, which are necessarily limits of type IV$_{h^{2,1}}$. 

\bibliography{papers_Max}
\bibliographystyle{JHEP}

\end{document}